\documentclass[final]{siamltex}

\usepackage{amsmath}

\usepackage{algorithm}
\usepackage[noend]{algorithmic}

\usepackage[english]{babel}
\usepackage{hyperref}
\usepackage[pdftex]{graphicx}

\newtheorem{example}[theorem]{Example}

\def\refeq#1{eqn (\ref{eq:#1})}

\def\refthm#1{Theorem \ref{thm:#1}}
\def\refThm#1{Theorem \ref{thm:#1}}

\def\refdef#1{Definition \ref{def:#1}}

\def\refexa#1{Example \ref{exa:#1}}

\def\reffig#1{Fig. \ref{fig:#1}}

\def\reftab#1{Table \ref{tab:#1}}
\def\refTab#1{Table \ref{tab:#1}}
\def\refalg#1{Algorithm \ref{alg:#1}}
\def\refAlg#1{Algorithm \ref{alg:#1}}
\def\refalgline#1{line \ref{algline:#1}}
\def\refsec#1{Section \ref{sec:#1}}

\newcommand{\ud}{\, \mathrm{d}}
\newcommand{\N}{\mathbf{N}}

\newcommand{\R}{\mathbf{R}}
\newcommand{\C}{\mathbf{C}}
\newcommand{\GrpO}{\mathrm{O}}

\newcommand{\setsep}{\ |\ }
\newcommand{\argmin}{\mathrm{argmin \,}}
\newcommand{\Complexity}{\mathcal{O}}

\def\supp{\mathrm{supp} \,}
\def\rank{\mathrm{rank} \,}

\begin{document}

\title{A Geometric Approach to Matrix Ordering}

\author{B.~O.~Fagginger Auer\thanks{Department of Mathematics, Utrecht University, P.O. Box 80010, 3508 TA Utrecht, the Netherlands (\texttt{B.O.FaggingerAuer@uu.nl}).} \and R.~H.~Bisseling\thanks{Department of Mathematics, Utrecht University, P.O. Box 80010, 3508 TA Utrecht, the Netherlands (\texttt{R.H.Bisseling@uu.nl}).}}

\date{\today}

\maketitle

\begin{abstract}
We present a recursive way to partition hypergraphs which creates and exploits hypergraph geometry and is suitable for many-core parallel architectures.
Such partitionings are then used to bring sparse matrices in a recursive Bordered Block Diagonal form (for processor-oblivious parallel LU decomposition) or recursive Separated Block Diagonal form (for cache-oblivious sparse matrix--vector multiplication).
We show that the quality of the obtained partitionings and orderings is competitive by comparing obtained fill-in for LU decomposition with SuperLU (with better results for $8$ of the $28$ test matrices) and comparing cut sizes for sparse matrix--vector multiplication with Mondriaan (with better results for $4$ of the $12$ test matrices).
The main advantage of the new method is its speed: it is on average $21.6$ times faster than Mondriaan.
\end{abstract}



\section{Introduction} \label{sec:intro}

\begin{figure}[h]
\begin{center}
\begin{tabular}{c@{\hspace{1cm}}c@{\hspace{1cm}}c}
\includegraphics[width=3.6cm]{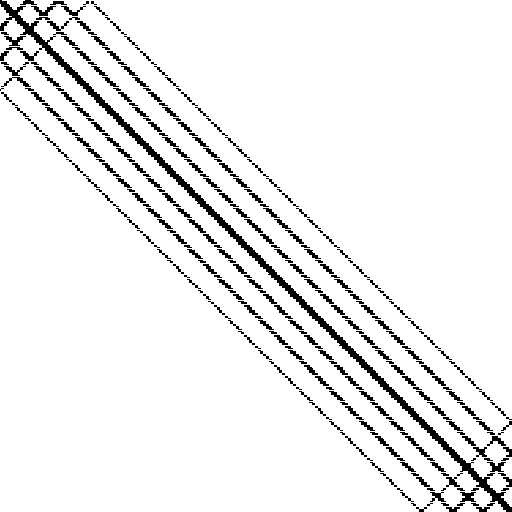} &
\includegraphics[width=3.5cm]{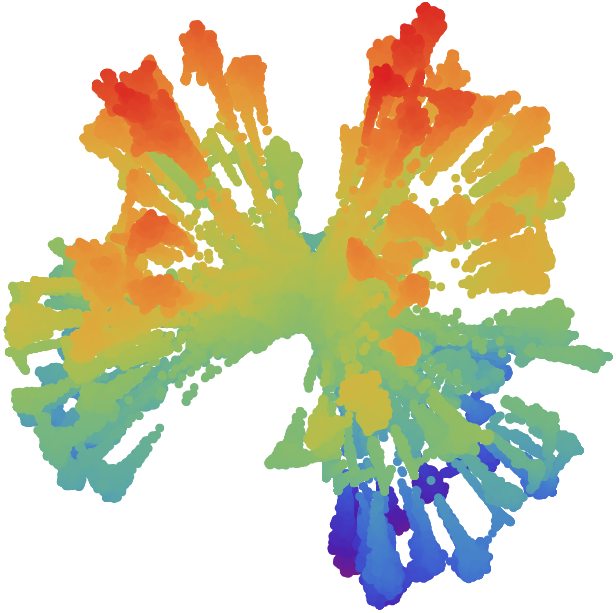} &
\includegraphics[width=3.6cm]{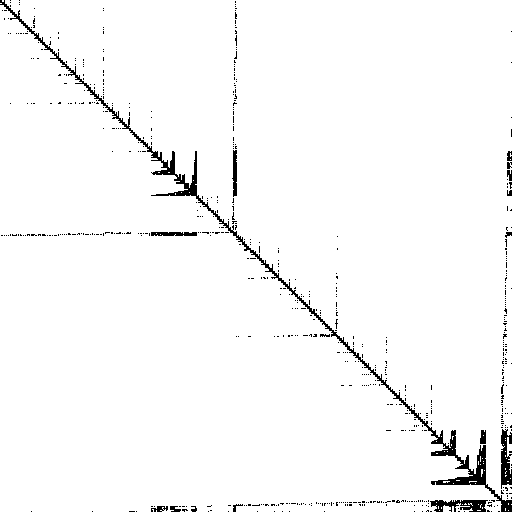} \\
\includegraphics[width=3.6cm]{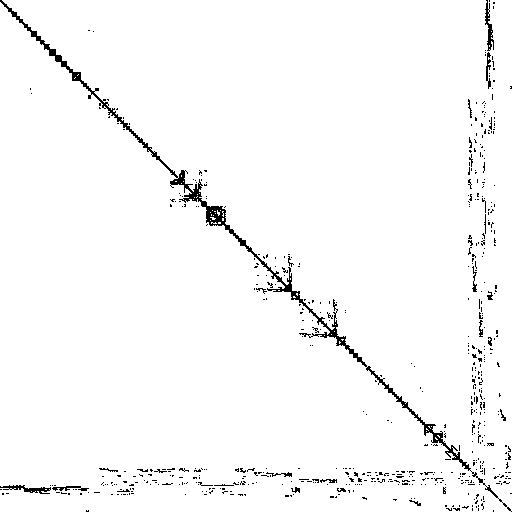} &
\includegraphics[height=3.4cm]{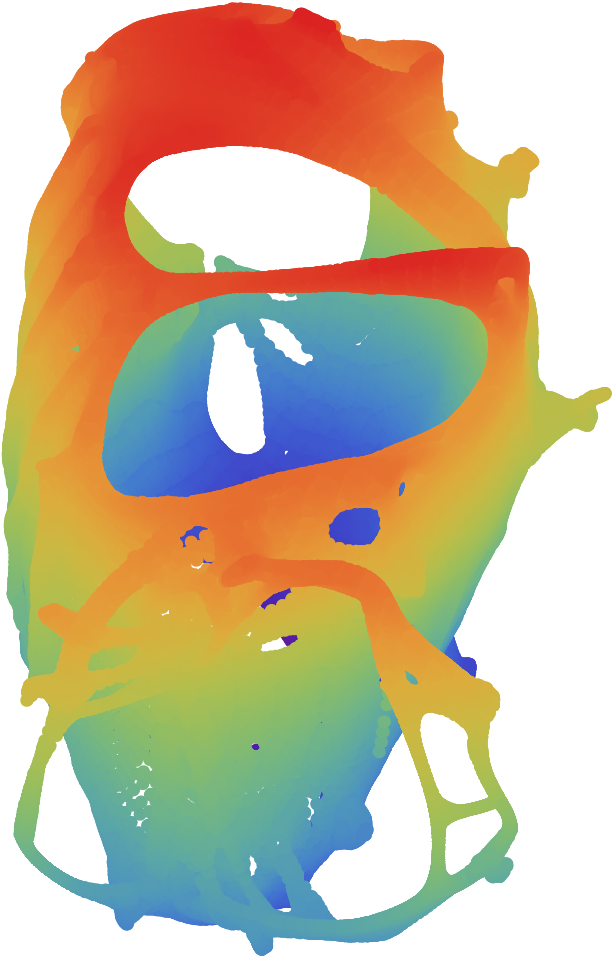} &
\includegraphics[width=3.6cm]{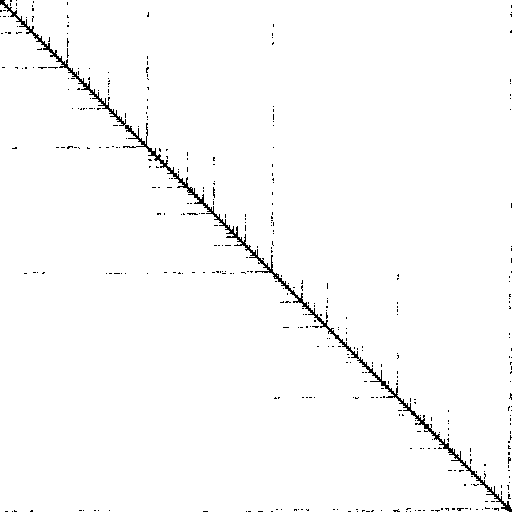}
\end{tabular}
\end{center}
\caption{The matrices \texttt{twotone} (top) and \texttt{ford2} (bottom). For the original matrix (left), a visual representation is created (middle), which in turn is used to permute the matrix to recursive Bordered Block Diagonal form (right).}
\label{fig:vmoprocess}
\end{figure}

With the increased development and availability of many-core processors (both CPUs and GPUs) it is important to have algorithms that can make use of these architectures.
To this end, we present a new recursive hypergraph partitioning algorithm that uses the underlying geometry of the hypergraph to generate the partitioning (which is largely done using shared-memory parallelism).
Hypergraph geometry may either be provided from the problem at hand or generated by the partitioning software.
This entire process is illustrated in \reffig{vmoprocess}.

\subsection{Hypergraphs} \label{sec:hypergraphs}

We will start with a brief introduction to hypergraphs (see \cite{Berge1976} for more information) and the ways in which they can be related to sparse matrices.

\begin{figure}
\begin{center}
\includegraphics[width=3.5cm]{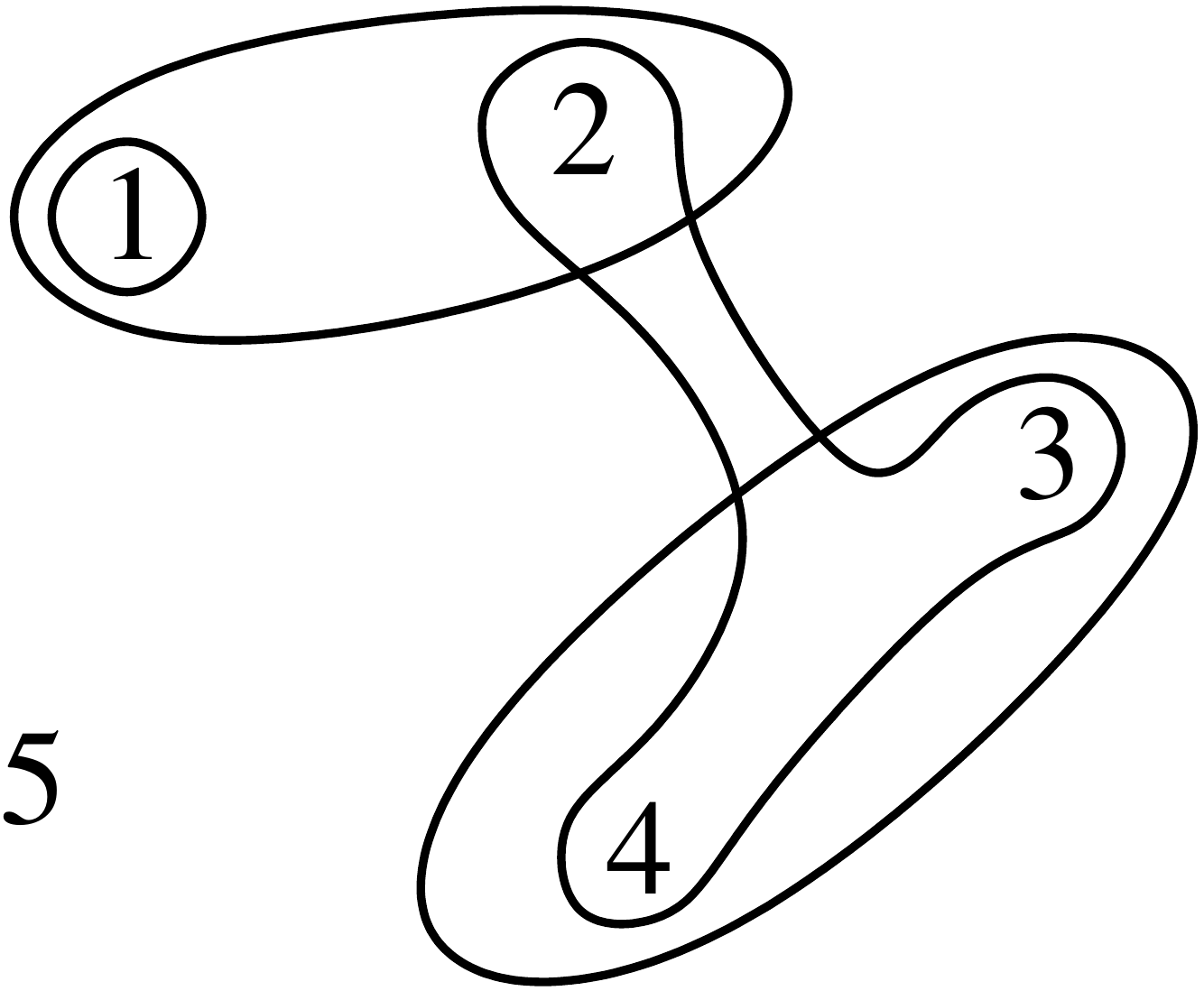}
\end{center}
\caption{Hypergraph $G = (V, E)$ with $V = \{1, 2, 3, 4, 5\}$ and $E = \{\{1\}, \{1, 2\}, \{2, 3, 4\}, \{3, 4\}\}$.}
\label{fig:hgexample}
\end{figure}

\begin{definition} \label{def:hypergraph}
A \emph{hypergraph} is a pair $G = (V, E)$ where $V$ is a set, the \emph{vertices} of the hypergraph $G$, and $E$ a collection of subsets of $V$ (so for all $e \in E$, $e \subseteq V$), the \emph{hyperedges} or \emph{nets} of $G$ (see \reffig{hgexample}).
We call a hypergraph $G = (V, E)$ \emph{weighted} when it is paired with functions $w : V \rightarrow [0, \infty[$ and $c : E \rightarrow [0, \infty[$ which assign \emph{weights} $w(v) \geq 0$ and \emph{costs} $c(e) \geq 0$ to vertices $v \in V$ and hyperedges $e \in E$, respectively.
We call a hypergraph $G = (V, E)$ simply a \emph{graph} if all hyperedges $e \in E$ are of the form $e = \{v, w\}$ with $v, w \in V$; in this case the graph is undirected and the hyperedges are called \emph{edges}.
We call a hypergraph $G = (V, E)$ \emph{finite} if $V$ is a finite set, in which case $|E| \leq 2^{|V|}$, so $E$ is finite as well.
\end{definition}

Hypergraphs possess a natural notion of \emph{duality}, where the roles played by the vertices and hyperedges are interchanged.

\begin{definition} \label{def:hypergraphdual}
Let $G = (V, E)$ be a hypergraph.

Then we define its \emph{dual hypergraph} as $G^* := (V^*, E^*)$ where $V^* := E$ and
$$E^* := \{ \{e \in E \setsep e \ni v \} \setsep v \in V\}.$$
If the hypergraph is weighted, we also exchange the vertex weights and hyperedge costs to make its dual a weighted hypergraph.
\end{definition}

The dual of the dual of a hypergraph is isomorphic to the original hypergraph, $(G^*)^* \cong G$.

\begin{table}
\begin{tabular}{|l|c|c|}
\hline
\textbf{Name} & $\mathbf{V}$ & $\mathbf{E}$ \\
\hline
Symmetric \cite{Parter1961} & $\{1, \ldots, m\}$ & $\{ \{i, j\} \setsep 1 \leq i \leq m, 1 \leq j \leq n, a_{i \, j} \neq 0\}$ \\
\hline
Bipartite \cite{Hendrickson2000} & $\{r_1, \ldots, r_m, c_1, \ldots, c_n\}$ & $\{ \{r_i, c_j\} \setsep 1 \leq i \leq m, 1 \leq j \leq n, a_{i \, j} \neq 0\}$ \\
\hline
Column-net \cite{Catalyurek1999} & $\{r_1, \ldots, r_m\}$ & $\{ \{r_i \setsep 1 \leq i \leq m, a_{i \, j} \neq 0\} \setsep 1 \leq j \leq n\}$ \\
\hline
Row-net \cite{Catalyurek1999} & $\{c_1, \ldots, c_n\}$ & $\{ \{c_j \setsep 1 \leq j \leq n, a_{i \, j} \neq 0\} \setsep 1 \leq i \leq m\}$ \\
\hline
Finegrain \cite{Catalyurek2001} & $\{v_{i \, j} \setsep a_{i \, j} \neq 0\}$ &
	$\underbrace{\{ \{v_{i \, j} \vert 1 \leq i \leq m, a_{i \, j} \neq 0\} \setsep 1 \leq j \leq n\}}_{\text{column hyperedges}}$ \\
	  & & 
	$\cup \underbrace{\{ \{v_{i \, j} \vert 1 \leq j \leq n, a_{i \, j} \neq 0\} \setsep 1 \leq i \leq m\}}_{\text{row hyperedges}}$ \\
\hline
\end{tabular}
\caption{Several common representations of an $m \times n$-matrix $A = \left( a_{i \, j} \right)$ by a hypergraph $G = (V, E)$.}
\label{tab:hypergraphmatrix}
\end{table}

One direct application of hypergraphs is to use them to represent sparse matrices, see \reftab{hypergraphmatrix}.
We can make a number of observations about these representations.
\begin{enumerate}
\item The symmetric representation is only sensible if the matrix is structurally symmetric, because only then we can recover the nonzero pattern of the original matrix from its representation.
\item The bipartite representation is a bipartite graph with the two parts consisting of the rows ($r_1, \ldots, r_m$) and the columns ($c_1, \ldots, c_n$) of the matrix.
\item The symmetric and bipartite representations are both undirected graphs instead of hypergraphs and the size of the symmetric representation is about half that of the bipartite representation.
\item The column-net and row-net representations are each other's dual.
\item The finegrain and bipartite representations are each other's dual.
This is also reflected in the fact that the hyperedges of the finegrain representation can be partitioned into two disjoint sets (the row and column hyperedges).
\end{enumerate}

We will make use of these observations in \refsec{lu}.

\subsection{Visual representations}

As stated in \refsec{intro}, we would like to exploit the underlying geometry of a hypergraph, usually representing a sparse matrix.

\begin{definition} \label{def:visualrepresentation}
Let $G = (V, E)$ be a given hypergraph.
Then a \emph{visual representation} of $G$ in $d \in \N$ dimensions is a mapping 
$$V \rightarrow \R^d$$
that reflects the structure of the underlying hypergraph.
\end{definition}

This definition is not very precise and therefore we will illustrate it by looking at a few examples.
Sometimes the visual representation of a matrix is directly available, for instance if the matrix is based on a triangulated mesh, such as the following \refexa{pothen}.
In other cases, the visual representation can be generated from the problem that the matrix represents, see \refexa{finiteelts}.
If no such information is available at all, we generate the visual representation ourselves, as discussed in \refsec{genvisrep}.

\begin{figure}
\begin{center}
\begin{tabular}{c@{\hspace{1cm}}c@{\hspace{1cm}}c}
\includegraphics[width=3.5cm]{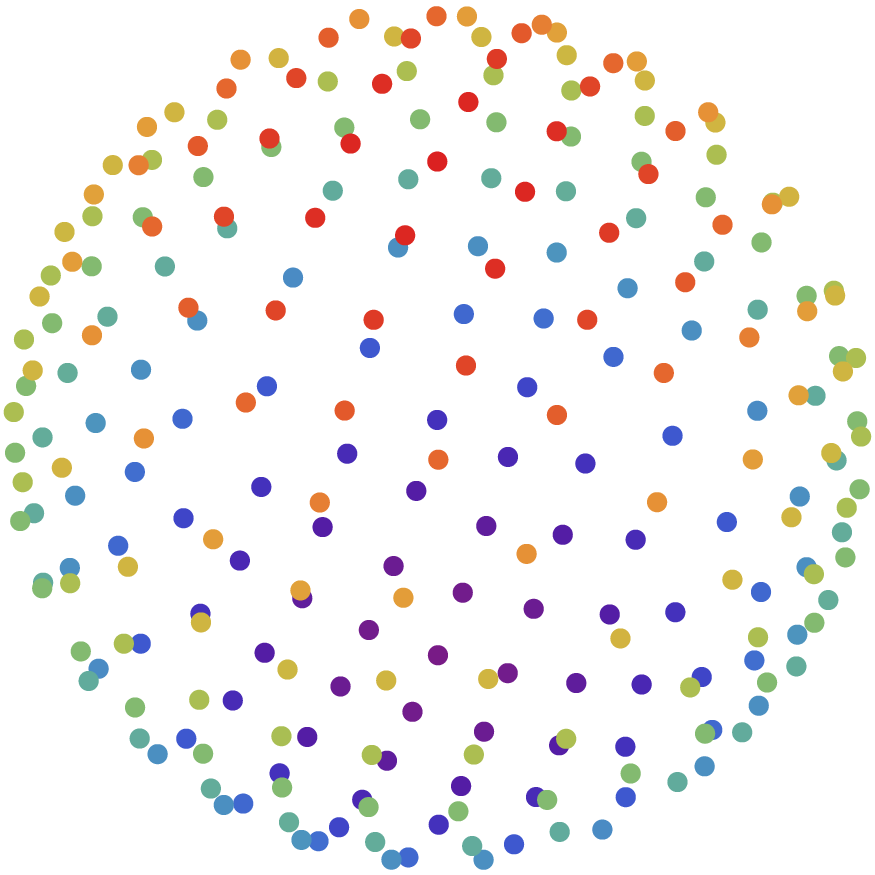} &
\includegraphics[width=3.5cm]{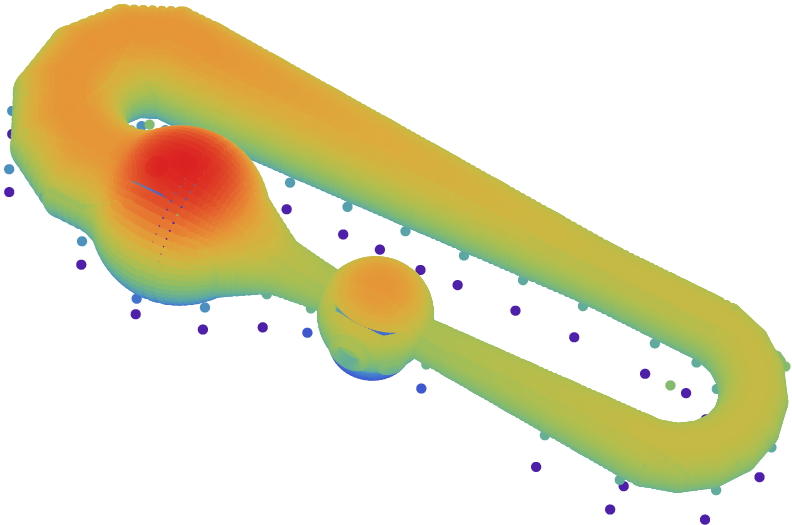} &
\includegraphics[width=3.5cm]{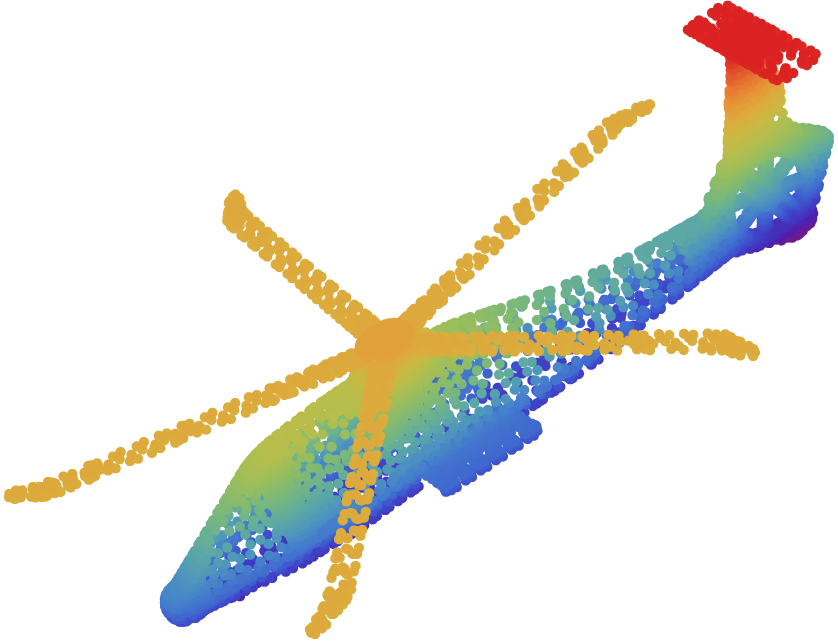} \\
\includegraphics[width=3.5cm]{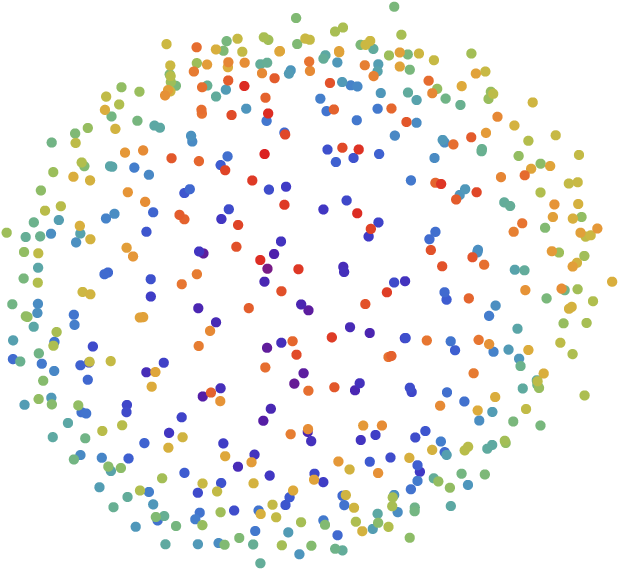} &
\includegraphics[width=3.5cm]{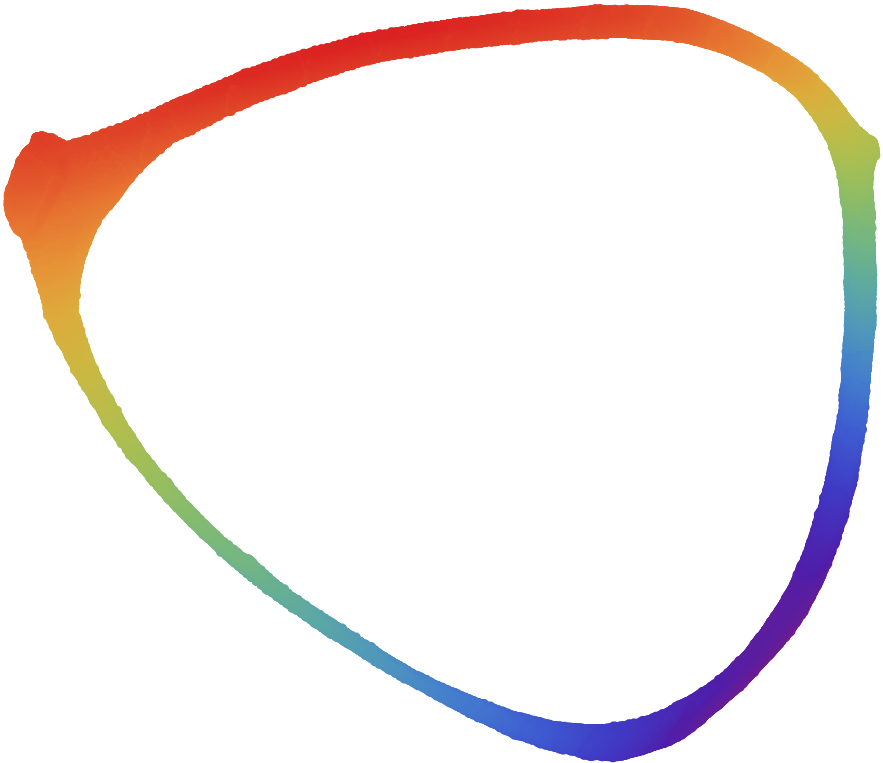} &
\includegraphics[width=3.5cm]{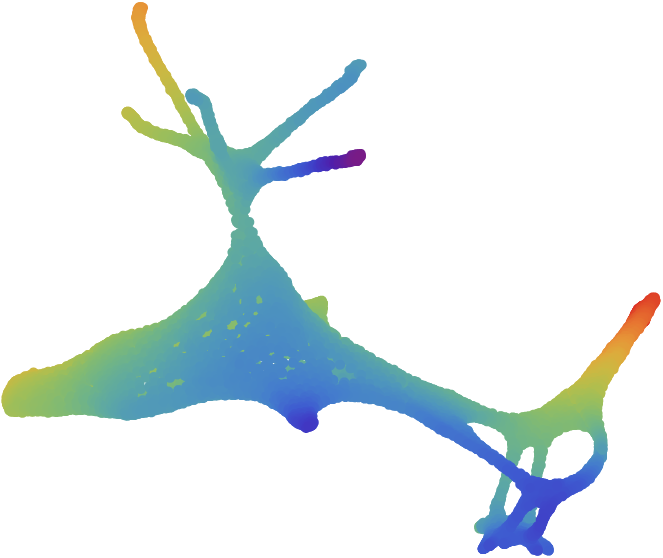}
\end{tabular}
\end{center}
\caption{From left to right: the matrices \texttt{sphere3}, \texttt{pwt}, and \texttt{commanche\_dual} with their original meshes (top, see \refexa{pothen}) and visual representations created using the techniques described in \refsec{genvisrep} (bottom, also compare with \cite{Hu2005}).}
\label{fig:pothenlayout}
\end{figure}

\begin{example} \label{exa:pothen}
The \texttt{pothen} collection of matrices, available from \cite{Davis2010}, consists of NASA structural engineering matrices collected by A.~Pothen.
We will take a closer look at the square pattern\footnote{Nonzero entries have numerical value $1$.} matrices from this collection, which have a natural visual representation.
Each row/column of these matrices corresponds to a vertex (the coordinates of which are supplied in a separate file) and each nonzero to an edge between the vertices corresponding to the row and column to which the nonzero belongs.
We can take a look at the matrices and their corresponding visual representation by plotting these vertices, as is done in \reffig{pothenlayout}.
These vertices give a visual representation of the symmetric hypergraph representation of the sparse matrix.
\end{example}

\begin{example} \label{exa:finiteelts}
Another important example of matrices with a natural visual representation are those arising in finite-element methods.
In this example, we will consider the Laplace equation on a compact smooth $d$-dimensional closed submanifold $\Omega \subseteq \R^d$ with compact smooth boundary $\partial \Omega$.
Let 
$$V := \{f \in C^{\infty}(\Omega, \R) \setsep f,\, \|\nabla f\| \in L^2(\Omega, \R), \ f \vert_{\partial \Omega} = 0 \}$$
be the collection of all smooth real-valued functions on $\Omega$ that are square-integrable, have square-integrable derivative, and vanish on the boundary $\partial \Omega$. Here the inner product is given by
$$\smash{\langle u, v \rangle = \int_{\Omega} u(x) \, v(x) \ud x + \int_{\Omega} \nabla u(x) \cdot \nabla v(x) \ud x.}$$
Let $g \in V$ be given.
We consider the problem of finding an $f \in V$ such that
\begin{equation} \label{eq:laplace}
\Delta f(x) = -g(x), \qquad (x \in \Omega).
\end{equation}
The first step in the finite-element method is to rewrite the above problem into its \emph{weak formulation}.\footnote{If the weak formulation satisfies the conditions of the Lax--Milgram theorem, the solution $f$ to the weak formulation is unique and hence also the solution to the original problem, provided such a solution exists \cite{Johnson1987}.}
Let $h \in V$ and suppose $f$ is a solution to \refeq{laplace}, then using integration by parts
$$
\int_{\Omega} g(x) \, h(x) \ud x = -\int_{\Omega} \Delta f(x) \, h(x) \ud x 
= -0 + \int_{\Omega} \nabla f(x) \cdot \nabla h(x) \ud x.
$$
Hence, every solution $f$ necessarily satisfies the weak formulation of this problem:
\begin{equation} \label{eq:laplaceweak}
a(f, h) = b(h), \qquad (h \in V),
\end{equation}
where
$$a(u, v) := \int_{\Omega} \nabla u(x) \cdot \nabla v(x) \ud x, \qquad b(u) := \int_{\Omega} g(x) \, u(x) \ud x.$$

If we now choose functions $h_1, \ldots, h_m \in V$, we can look at the approximate solution $\phi$ to \refeq{laplaceweak} in the subspace $V_m := \langle h_1, \ldots, h_m \rangle_{\R} \subseteq V$ spanned by $h_1, \ldots, h_m$.
Because $\phi \in V_m$, there exist coefficients $\phi_1, \ldots, \phi_m \in \R$ such that $\phi = \sum_{i = 1}^m \phi_i \, h_i$, and therefore,
$$b(h_j) = a(\phi, h_j) = a \left( \sum_{i = 1}^m \phi_i \, h_i, h_j \right)
 = \sum_{i = 1}^m \phi_i \, a(h_i, h_j), \qquad (1 \leq j \leq m).$$
So solving \refeq{laplaceweak} in $V_m$ for $\phi$ amounts to solving the linear system
$$\sum_{i = 1}^m a(h_i, h_j) \, \phi_i = b(h_j), \qquad (1 \leq j \leq m),$$
for $(\phi_1, \ldots, \phi_m) \in \R^m$.
By choosing $h_1, \ldots, h_m$ such that their supports only have a small overlap we create a sparse matrix $A \in \R^{m \times m}$ with entries $a_{i \, j} := a(h_j, h_i)$ that are nonzero only for the $1 \leq i, j \leq m$ where $\supp h_i \cap \supp h_j \neq \emptyset$.
This can be done, for instance, by triangulating $\Omega$ and choosing for each vertex $i$ in the triangulation a function $h_i$ with support contained in all triangles adjacent to the vertex $i$.

We can now directly create a visual representation for the various hypergraph representations of the matrix $A$ (\reftab{hypergraphmatrix}):
\begin{enumerate}
\item for the symmetric, bipartite, column-net, and row-net representations, we map each vertex $i$, $r_i$, and $c_i$ to the center of $\supp h_i$ in $\R^d$,
\item for the finegrain representation, we map each vertex $v_{i \, j}$ to the center of $\supp h_i \cap \supp h_j$ in $\R^d$.
\end{enumerate}
Thus, in this case, we can generate a visual representation from our original problem with little extra effort.
\end{example}

However, not all matrices have an immediate geometric origin (e.g. \texttt{twotone}), so we would like to be able to create such a visual representation ourselves if it cannot be provided directly.
Hence, we will continue by discussing how to create such a visual representation in \refsec{genvisrep}.
In \refsec{partitioning}, we show how to use visual representations for hypergraph partitioning, and in \refsec{lu} we apply this in the context of sparse LU decomposition.
We conclude by implementing these methods and comparing them with both SuperLU and Mondriaan in \refsec{experiments}.

\section{Creating visual representations} \label{sec:genvisrep}

To create visual representations, we employ the method described in \cite{Fruchterman1991, Hu2005, Walshaw2003}, generalized to visualizing hypergraphs: we let the vertices of the hypergraph repel each other by an electrostatic-like force and let the hyperedges bind their vertices together like rubber bands.
We model this as the minimization of an energy function where we lay out the graph in at least three dimensions (to prevent having to treat the special cases $d = 1$ and $d = 2$ where the form we propose for the energy function is not appropriate).
For brevity, we will simply enumerate the vertices of the hypergraph we consider, thus assuming that $V = \{1, \ldots, k\}$ for some $k \in \N$.

\begin{definition} \label{def:graphenergy}
Let $G = (\{1, \ldots, k\}, \{e_1, \ldots, e_l\})$ be a finite weighted hypergraph with vertex weights $w_1, \ldots, w_k > 0$, hyperedge costs $c_1, \ldots, c_l > 0$, and let $d \geq 3$ be the dimension of the target space.
Define
$$U := \{(x_1, \ldots, x_k) \in \R^{d \times k} \setsep \forall 1 \leq i < j \leq k : x_i \neq x_j\}.$$

Then the \emph{energy function of $G$} is defined as
\begin{equation} \label{eq:graphenergy}
f(x_1, \ldots, x_k) := \underbrace{\frac{\alpha}{2} \sum_{j = 1}^l c_j \sum_{i \in e_j} \|x_i - z_j\|^{\gamma}}_{\text{rubber bands}} + \underbrace{\frac{\beta}{2} \sum_{j = 1}^k \sum_{\substack{i = 1 \\ i \neq j}}^k \frac{w_i \, w_j}{\|x_i - x_j\|^{\delta}}}_{\text{repelling charges}},
\end{equation}
for $(x_1, \ldots, x_k) \in U$, where $\alpha, \beta, \gamma, \delta > 0$ are constants, and for $1 \leq j \leq l$ the center of hyperedge $e_j$ is defined as
$$\smash[t]{z_j := \frac{1}{|e_j|} \sum_{i \in e_j} x_i.}$$
\end{definition}

Now, we will generate a visual representation by finding
\begin{equation} \label{eq:mingraphenergy}
\argmin \{f(x_1, \ldots, x_k) \setsep (x_1, \ldots, x_k) \in U\},
\end{equation}
where $f$ is the energy function of our hypergraph $G$.
Approximate solutions to \refeq{mingraphenergy} generated by our algorithm are shown in \reffig{pothenlayout}.

\subsection{Constants}

We can tinker with $f$ by varying the constants $\alpha$, $\beta$, $\gamma$, and $\delta$.
First, we can make a number of observations about $U$, $f$, and their symmetries.
\begin{enumerate}
\item $\{(R(x_1) + x, \ldots, R(x_k) + x) \setsep (x_1, \ldots, x_k) \in U\} = U$ for all $x \in \R^d$ and all orthogonal transformations $R \in \GrpO(\R^d)$.
\item $U$ is an open subset of $\R^{d \times k}$ and $U = \theta \, U$ for all $\theta \in \R \setminus \{0\}$.
\item $f(R(x_1) + x, \ldots, R(x_k) + x) = f(x_1, \ldots, x_k)$ for all $x \in \R^d$, $R \in \GrpO(\R^d)$.
\item $f \in C(U, [0, \infty[\;)$ is continuous; $f$ is always bounded from below by $0$; and if $k \geq 2$, $f$ is unbounded from above.
\end{enumerate}

So $U$ and $f$ are invariant under all translations and orthogonal transformations in $\R^d$, but while $U$ is scaling-invariant, $f$ in general is not.

Let $\theta > 0$ and $(x_1, \ldots, x_k) \in U$.
Then
$$f(\theta \, x_1, \ldots, \theta \, x_k) = \frac{\alpha \, \theta^{\gamma}}{2} \sum_{j = 1}^l c_j \sum_{i \in e_j} \|x_i - z_j\|^{\gamma} + \frac{\beta}{2 \, \theta^{\delta}} \sum_{j = 1}^k \sum_{\substack{i = 1 \\ i \neq j}}^k w_i \, w_j \, \|x_i - x_j\|^{-\delta}.$$
So in particular, for all $(x_1, \ldots, x_k) \in U$ we have that
$$\frac{1}{\beta} \left( \frac{\beta}{\alpha} \right)^{\frac{\delta}{\gamma + \delta}} \, f \left( \left( \frac{\beta}{\alpha} \right)^{\frac{1}{\gamma + \delta}} \, x_1, \ldots, \left( \frac{\beta}{\alpha} \right)^{\frac{1}{\gamma + \delta}} \, x_k \right) = \tilde{f}(x_1, \ldots, x_k),$$
where $\tilde{f}$ is given by \refeq{graphenergy}, but with $\alpha = \beta = 1$.
Therefore, we can pick $\alpha = \beta = 1$ without loss of generality; this just scales the minimum of $f$ by an overall factor (a similar scaling property is derived in \cite[Theorem 1]{Hu2005}).

The function $f$ is continuously differentiable for $\gamma \geq 2$.
Calculating the partial derivatives of $f$, we find that for $1 \leq m \leq d$, $1 \leq n \leq k$,
\begin{multline} \label{eq:graphenergygradient}
\frac{\partial f(x_1, \ldots, x_k)}{\partial x_{m \, n}} =
- \frac{\alpha \, \gamma}{2} \sum_{\{j \setsep n \in e_j\}} \frac{c_j}{|e_j|} \sum_{i \in e_j} \|x_i - z_j\|^{\gamma - 2} \, (x_{m \, i} - z_{m \, j}) \\
+ \frac{\alpha \, \gamma}{2} \sum_{\{j \setsep n \in e_j\}} c_j \, \|x_n - z_j\|^{\gamma - 2} \, (x_{m \, n} - z_{m \, j})
+ \underbrace{\beta \, \delta \sum_{\substack{i = 1 \\ i \neq n}}^k \frac{w_i \, w_n}{\|x_i - x_n\|^{\delta + 2}} \, (x_{m \, i} - x_{m \, n})}_{\text{repelling charges}}.
\end{multline}
Note that \refeq{graphenergygradient} simplifies considerably if we pick $\gamma = 2$, as $\sum_{i \in e_j} \|x_i - z_j\|^{2 - 2} \, (x_{m \, i} - z_{m \, j}) = (\sum_{i \in e_j} x_{m \, i}) - |e_j| \, z_{m \, j} = 0$, which makes the first term disappear, and ensures that $f \in C^{\infty}(U, [0, \infty[)$ is smooth.
This motivates us to pick $\gamma = 2$.

Calculating the repelling-charges part of \refeq{graphenergygradient} for $n = 1, \ldots, k$ requires $\Complexity(k^2)$ evaluations which is too expensive for the hypergraphs we envision, with a huge number of vertices $k$.
Luckily, we can circumvent this problem using techniques from \cite{Hu2005} and \cite{Nakasato2009}:
we will build the $d$-dimensional equivalent of an octree to group the repelling charges into clusters and treat far-away clusters of charges as a single, but heavier charge.
To ensure that this works properly the location of this larger charge will be the weighted average location of the cluster and the weight of the charge will be set equal to the sum of the charge weights in the cluster.
Note that the repelling-charges part of $f$ in \refeq{graphenergy} consists of applications of the map $x \mapsto \|x\|^{-\delta}$ for $x \in \R^d$, the Laplacian of which is given by $x \mapsto \delta \, (\delta - (d - 2)) \, \|x\|^{-\delta - 2}$.
Hence, we can ensure that this part of $f$ is harmonic (i.e. with vanishing Laplacian) by choosing $\delta = d - 2$.
This will in turn make our energy function behave well when treating clusters of far-away charges as a single, heavier charge, because of the \emph{mean-value property} of harmonic functions (see \cite[Theorem 1.6 and 1.24]{Axler1992}).

Recapitulating the above:
\begin{enumerate}
\item we can pick $\alpha = \beta = 1$ since this only scales a solution to \refeq{mingraphenergy},
\item we should pick $\gamma = 2$ to be able to calculate the rubber band contribution to \refeq{graphenergygradient} efficiently,
\item we should pick $\delta = d - 2$ to be able to approximate the repelling charge contribution to \refeq{graphenergygradient} by treating clusters of charges as a single, heavier charge.
\end{enumerate}

Inserting these constants we obtain the following expressions for $f$ and its partial derivatives:
\begin{align}
f(x_1, \ldots, x_k)
& = \frac{1}{2} \sum_{j = 1}^l c_j \sum_{i \in e_j} \|x_i - z_j\|^2
+ \frac{1}{2} \sum_{j = 1}^k \sum_{\substack{i = 1 \\ i \neq j}}^k \frac{w_i \, w_j}{\|x_i - x_j\|^{d - 2}}, \label{eq:graphenergysimple} \\
\frac{\partial f(x_1, \ldots, x_k)}{\partial x_{m \, n}}
& = \sum_{\{j \setsep n \in e_j\}} c_j \, (x_{m \, n} - z_{m \, j})
+ (d - 2) \sum_{\substack{i = 1 \\ i \neq n}}^k \frac{w_i \, w_n}{\|x_i - x_n\|^d} \, (x_{m \, i} - x_{m \, n}). \label{eq:graphenergygradientsimple}
\end{align}

\subsection{Connectedness}

Before we start describing an algorithm to solve \refeq{graphenergy}, we should take care to ensure that such a solution actually exists.

\begin{theorem} \label{thm:disconnectedminimum}
Let $G = (V, E)$, $U$, and $f$ be given as in \refdef{graphenergy} and suppose $G$ is non-empty.

Then precisely one of the following statements is true:
\begin{enumerate}
\item there exists a solution $(x_1, \ldots, x_k) \in U$ to \refeq{mingraphenergy},
\item $G$ is disconnected: we can write $V = V_1 \cup V_2$ as a disjoint union with $V_1, V_2 \neq \emptyset$, such that for all $e \in E$ we have $e \subseteq V_1$ or $e \subseteq V_2$.
\end{enumerate}
\end{theorem}
\begin{proof}
Suppose $G$ is disconnected.
Without loss of generality, we can order the vertices such that $V_1 = \{1, \ldots, k'\}$ and $V_2 = \{k' + 1, \ldots, k\}$.
Suppose $(x_1, \ldots, x_k) \in U$ has minimum energy.
Then for all $y \in \R^d \setminus \{0\}$ we have
\begin{multline*}
f(x_1 + y, \ldots, x_{k'} + y, x_{k'} - y, \ldots, x_k - y) \\
= f(x_1, \ldots, x_k)
+ \beta \sum_{i = 1}^{k'} \sum_{j = k' + 1}^k \left( \frac{w_i \, w_j}{\|2 \, y + x_i - x_j\|^{\delta}} - \frac{w_i \, w_j}{\|x_i - x_j\|^{\delta}} \right).
\end{multline*}
This equation holds because all hyperedges $e_j \in E$ are either completely contained in $V_1$ or completely contained in $V_2$, such that all $x_i$ with $i \in e_j$ are either translated by $+y$ or $-y$, respectively.
This ensures that for all hyperedges $e_j \in E$ and vertices $i \in e_j$, the difference $x_i - z_j$ remains the same, which in turn leaves the first term of \refeq{graphenergy} unchanged.
For the second term of \refeq{graphenergy}, we find that the difference $x_i - x_j$ only changes if $i \in V_1$ and $j \in V_2$, or vice versa.

For all $1 \leq i \leq k'$, $k' < j \leq k$, we have
$$\lim_{r \rightarrow \infty} \frac{w_i \, w_j}{\|2 \, (r \, y) + x_i - x_j\|^{\delta}} 
= \lim_{r \rightarrow \infty} \frac{1}{|r|^{\delta}} \, \frac{w_i \, w_j}{\|2 \, y + (x_i - x_j)/r\|^{\delta}}
= 0$$
as $y \neq 0$ and $\delta > 0$.
In particular there exists an $r > 0$ such that for all $1 \leq i \leq k'$, $k' < j \leq k$ we have
$$\smash{\frac{w_i \, w_j}{\|2 \, (r \, y) + x_i - x_j\|^{\delta}} - \frac{w_i \, w_j}{\|x_i - x_j\|^{\delta}} < 0.}$$
So for this $r$ we have
$$f(x_1 + r \, y, \ldots, x_{k'} + r \, y, x_{k' + 1} - r \, y, \ldots, x_k - r \, y) < f(x_1, \ldots, x_k),$$
hence $(x_1, \ldots, x_k)$ does not have minimum energy; we have reached a contradiction.
Therefore, no solution to \refeq{mingraphenergy} exists.

Suppose conversely that $G$ has no disconnected components: for any disjoint union $V = V_1 \cup V_2$ with $V_1, V_2 \neq \emptyset$ there exists an $e \in E$ such that $e \cap V_1 \neq \emptyset$ and $e \cap V_2 \neq \emptyset$.
So in particular for every $v, w \in V$ there exists a path of hyperedges $e_1, \ldots, e_n \in E$ such that $v \in e_1$, $w \in e_n$, and $e_{j - 1} \cap e_j \neq \emptyset$ for all $1 < j \leq n$.
We can see this by writing $V$ as the disjoint union $\{v\} \cup (V \setminus \{v\})$ which gives us $e_1$ and continuing by induction to obtain $e_{j + 1}$ from the disjoint union $V = (e_1 \cup \ldots \cup e_j) \cup (V \setminus (e_1 \cup \ldots \cup e_j))$ until we reach $w$ (which will happen eventually as each new hyperedge adds at least one new vertex and our hypergraph is finite).

Suppose that for a given $R > 0$, $(x_1, \ldots, x_k) \in U$ satisfies
\begin{equation} \label{eq:energymaxradius}
\max_{1 \leq i < j \leq k} \|x_i - x_j\| \geq R.
\end{equation}
We will now focus on two vertices for which the relative distance in \refeq{energymaxradius} is maximal.
As described above there exists a path of hyperedges between these two vertices.
So there exist vertices $i_1, \ldots, i_{n + 1} \in V$ and edges $e_1, \ldots, e_n \in E$ such that $\|x_{i_1} - x_{i_{n + 1}}\|$ is maximal, $i_1 \in e_1$, $i_{n + 1} \in e_n$, and $i_j \in e_{j - 1} \cap e_j$ for $1 < j \leq n$.
Along this path, we have
\begin{align*}
R & \leq \|x_{i_1} - x_{i_{n + 1}}\|
 = \left\|x_{i_1} - z_1 + \sum_{m = 2}^n (z_{j - 1} - x_{i_m} + x_{i_m} - z_j) + z_n - x_{i_{n + 1}} \right\| \\
& \leq \|x_{i_1} - z_1\| + \sum_{m = 2}^n (\|x_{i_m} - z_{j - 1}\| + \|x_{i_m} - z_j\|) + \|x_{i_{n + 1}} - z_n\|.
\end{align*}
Hence, one of these $2 \, n$ terms must be at least $R / (2 \, n)$.
So there exist a vertex $p \in V$ and a hyperedge $e_q \in E$ such that $p \in e_q$ and $\|x_p - z_q\| \geq R/(2 \, n) \geq R/(2 \, l).$
Therefore, $f$ satisfies
\begin{equation} \label{eq:energyblowupmax}
f(x_1, \ldots, x_k)
\geq \frac{\alpha}{2} \, c_q \, \|x_p - z_q\|^{\gamma}
\geq \frac{\alpha}{2 \, (2 \, l)^{\gamma}} \, \left( \min_{1 \leq j \leq l} c_j \right) \, R^{\gamma}.
\end{equation}

Suppose that for a given $r > 0$, $(x_1, \ldots, x_k) \in U$ satisfies
$$\min_{1 \leq i < j \leq k} \|x_i - x_j\| \leq r.$$
Fix $1 \leq i < j \leq k$ such that $\|x_i - x_j\|$ is minimal, then
\begin{equation} \label{eq:energyblowupmin}
f(x_1, \ldots, x_k) \geq \beta \, \frac{w_i \, w_j}{\|x_i - x_j\|^{\delta}} \geq \beta \, w_i \, w_j \, \frac{1}{r^{\delta}} \geq \beta \, \left( \min_{1 \leq i \leq k} w_i^2 \right) \, \frac{1}{r^{\delta}}.
\end{equation}

Note that $f$ is invariant under translations, so that without loss of generality we can restrict ourselves to solutions with $x_1 = 0$.
Let
$$E := f((0, 0, \ldots, 0), (1, 0, \ldots, 0), (2, 0, \ldots, 0), \ldots, (k - 1, 0, \ldots, 0))$$
be an upper bound for the minimum value of $f$.

By \refeq{energyblowupmax} we know that there exists an $R > 0$ such that $f(0, x_2, \ldots, x_k) > E$ whenever $\max_{i \neq j} \|x_i - x_j\| > R$.
On the other hand, by \refeq{energyblowupmin} there exists an $r > 0$ such that $f(0, x_2, \ldots, x_k) > E$ whenever $\min_{i \neq j} \|x_i - x_j\| < r$.
So if $(0, x_2, \ldots, x_k) \in U$ has minimal energy, then necessarily $(0, x_2, \ldots, x_k) \in C$ where
$$C := \bigcap_{1 \leq i < j \leq k} \left\{ (x_1 = 0, x_2, \ldots, x_k) \in \R^{d \times k} \setsep r \leq \|x_i - x_j\| \leq R \right\}.$$
As $C \subseteq \R^{d \times k}$ is both closed and bounded (because $x_1 = 0$), $C$ is compact by Theorem 1.8.17 in \cite{Duistermaat2004i}.
We also have that $C \subseteq U$ and $f$ is continuous on $U$, so $f$ is continuous on the compact set $C$ and therefore $f \vert_C$ attains its minimum at a certain point in $C$ by Theorem 1.8.8 of \cite{Duistermaat2004i}.
Since the minimum of $f$ necessarily lays within $C$ by construction, this is a solution to \refeq{mingraphenergy}.
\end{proof}

Therefore, in solving \refeq{mingraphenergy}, we should restrict ourselves to connected hypergraphs.
If the provided hypergraph is not connected, we have to treat each connected component separately, where a solution is guaranteed to exist by \refthm{disconnectedminimum}.

\subsection{Algorithm}

\begin{algorithm}
\begin{algorithmic}[1]
\FORALL{$v \in V$}
	\STATE $p_v \gets v$;
\ENDFOR
\FORALL{$e \in E$}
	\FORALL{$v \in e$}
		\WHILE{$p_v \neq p_{p_v}$}
			\STATE $p_v \gets p_{p_v}$;
		\ENDWHILE
	\ENDFOR
	\STATE $p \gets \min \{p_v \setsep v \in e\}$;
	\FORALL{$v \in e$}
		\STATE $p_{p_v} \gets p$;
	\ENDFOR
\ENDFOR
\FORALL{$v \in V$}
	\WHILE{$p_v \neq p_{p_v}$}
		\STATE $p_v \gets p_{p_v}$;
	\ENDWHILE
\ENDFOR
\end{algorithmic}
\caption{Determines the connected components of a hypergraph $G = (V, E)$, with $V = \{1, \ldots, k\}$.
Vertices $v$ and $w$ belong to the same connected component of $G$ iff $p_v = p_w$.
This algorithm is adapted from \cite[Section 21.3]{Cormen2009}.}
\label{alg:connectedcomponents}
\end{algorithm}

We use the above observations to create a multilevel algorithm which approximates a minimum-energy solution.
Firstly, we find the connected components via \refalg{connectedcomponents}.
This algorithm works by creating a forest, i.e. a collection of rooted trees, where the parent of a vertex $v \in V$ is denoted by $p_v \in V$, and each root satisfies $p_v = v$.
The algorithm merges the trees of all vertices contained in a hyperedge, for all hyperedges.
After this is completed, each vertex is directly attached to its root, which represents its connected component.

\begin{algorithm}
\begin{algorithmic}[1]
\STATE $G^0 \gets G$; $j \gets 0$;
\WHILE{$G^j$ is too large}
	\STATE coarsen $G^j$ to $G^{j + 1}$ with surjective $\pi^j : V_{G^j} \rightarrow V_{G^{j + 1}}$;
	\STATE $j \gets j + 1$;
\ENDWHILE
\STATE let $x^j$ be a random visual representation for $G^j$
\STATE (so pick $x^j(v) \in \R^d$ at random for all vertices $v \in V_{G^j}$);
\WHILE{$j \geq 0$}
	\STATE improve $x^j$ by \refalg{vmominenergy}; \label{algline:vmovisrepimp}
	\IF{$j > 0$}
		\FORALL{vertices $v \in V_{G^{j - 1}}$ \textbf{parallel}} \label{algline:vmovisreppar}
			\STATE $x^{j - 1}(v) \gets $ scale $x^j(\pi^j(v))\, + $ small random displacement;
		\ENDFOR
	\ENDIF
	\STATE $j \gets j - 1$;
\ENDWHILE
\end{algorithmic}
\caption{Finds solutions to \refeq{mingraphenergy} for a given connected hypergraph $G = (V, E)$, adopted from \cite{Hu2005}.}
\label{alg:vmovisrep}
\end{algorithm}

Secondly, we apply \refalg{vmovisrep} to each connected component to generate a visual representation.
We create a hierarchy of coarsenings of the hypergraph and visual representations for these coarser hypergraphs, to avoid getting stuck in local energy minima.
Here, we follow the graph visualization algorithm from \cite{Hu2005}.
We coarsen hypergraphs by creating a maximal matching that is constructed greedily from heavy (high cost) hyperedges, as heavy hyperedges have the tendency to pull the vertices contained in them closer together when we try to find the energy minimum, and then merging the matched vertices.
To prevent star hypergraphs from forming (which disrupt this matching procedure \cite{Davis2010}) we also always match single-neighbor vertices to their neighbor.
Coarsening a hypergraph $G$ to a hypergraph $H$ in this way, we obtain a surjective map $\pi : V_G \rightarrow V_H$ which maps each collection of matched vertices of $G$ to a single vertex of $H$.
We also merge hyperedges $e, e' \in E_G$ satisfying $\pi(e) = \pi(e')$ (where $\pi(e) = \{\pi(v) \setsep v \in e\}$) to a single hyperedge and set the cost of this new hyperedge to the sum of the costs of $e$ and $e'$.
At \refalgline{vmovisreppar} of \refalg{vmovisrep}, we introduce the \textbf{parallel do} construct.
We use it to denote parallel for-loops that can directly be parallelized because their iterations are independent.

For the coarsest version of the hypergraph, we start out with a random visual representation, which is then improved by \refalg{vmominenergy} using the steepest descent method.
After the visual representation has been improved for the coarse hypergraph we scale it and add small random displacements, to ensure that different vertices in the fine hypergraph do not occupy the same position.
Thus, we obtain a visual representation for the fine hypergraph.
This layout is then again improved by \refalg{vmominenergy}.
We continue doing this until we obtain a visual representation for the original hypergraph.

\begin{algorithm}
\begin{algorithmic}[1]
\WHILE{we are not satisfied with the solution}
	\STATE build tree $T$ recursively clustering $x_1, \ldots, x_k$;
	\FOR{$n = 1$ \TO $k$ \textbf{parallel}}
		\STATE $y_n \gets 0$;
	\ENDFOR
	\FOR{$j = 1$ \TO $l$ \textbf{parallel}}
		\STATE $z_j \gets \frac{1}{|e_j|} \sum_{i \in e_j} x_i$;
	\ENDFOR
	\FOR{$n = 1$ \TO $k$ \textbf{parallel}}
		\FORALL{$j$ such that $n \in e_j$}
			\STATE $y_n \gets y_n + c_j \, (x_n - z_j)$; (cf. \refeq{graphenergygradientsimple})
		\ENDFOR
	\ENDFOR
	\FOR{$n = 1$ \TO $k$ \textbf{parallel}}
		\STATE $t \gets $ root of $T$;
		\WHILE{$t \neq \text{dummy}$}
			\IF{$x_n$ is far away from $t$}
				\STATE $y_n \gets y_n + (d - 2) \, w_t \, w_n \, \|x_t - x_n\|^{-d} \, (x_t - x_n)$; (cf. \refeq{graphenergygradientsimple})
				\STATE $t \gets $ sibling of $t$;
			\ELSIF{$t$ has a child}
				\STATE $t \gets $ child of $t$;
			\ELSE
				\FORALL{$i \in t$, $i \neq n$}
					\STATE $y_n \gets y_n + (d - 2) \, w_i \, w_n \, \|x_i - x_n\|^{-d} \, (x_i - x_n)$; (cf. \refeq{graphenergygradientsimple})
				\ENDFOR
				\STATE $t \gets $ sibling of $t$;
			\ENDIF
		\ENDWHILE
	\ENDFOR
	\STATE determine appropriate stepsize $\alpha > 0$;
	\FOR{$n = 1$ \TO $k$ \textbf{parallel}}
		\STATE $x_n \gets x_n - \alpha \, y_n$;
	\ENDFOR
\ENDWHILE
\end{algorithmic}
\caption{Improves a given approximate solution $(x_1, \ldots, x_k) \in U$ to \refeq{mingraphenergy} for a hypergraph $G = (\{1, \ldots, k\}, \{e_1, \ldots, e_l\})$ using steepest descent.}
\label{alg:vmominenergy}
\end{algorithm}

\refAlg{vmominenergy} gives the details of the improvement procedure on \refalgline{vmovisrepimp} of \refalg{vmovisrep}.
The part of the gradient of the energy function $f$ belonging to $x_n$, is denoted by $y_n$.
We create a tree $T$ which recursively groups the points belonging to the visual layout; this tree consists of nodes $t \in T$ which have position $x_t \in \R^d$ (the average position of all points contained in $t$) and weight $w_t > 0$ (the sum of the weights of all vertices contained in $t$).
The functionality of siblings of nodes in the tree is extended by letting the root have a $\text{dummy}$ node as sibling (denoting the end of the tree traversal), and letting nodes without siblings have the sibling of their parent as sibling.
This facilitates a fast, direct tree traversal without backtracking \cite{Nakasato2009}.
During steepest descent, we determine the step size by comparing the bounding box volume to the individual gradients of the vertices to prevent blowup for the first few steps and then decrease the found stepsize by multiplying it by $0.9$ \cite{Hu2005}.
Note that apart from the building of the tree grouping the hypergraph vertices, almost all parts of \refalg{vmominenergy} consist of $d$-dimensional floating point arithmetic that is directly parallellizable over all vertices and hyperedges.
This makes \refalg{vmominenergy} suitable for many-core architectures such as GPUs.

\section{Partitioning} \label{sec:partitioning}

Suppose that $x_1, \ldots, x_k \in \R^d$ is a visual representation of our hypergraph $G = (\{1, \ldots, k\}, \{e_1, \ldots, e_l\})$, obtained either directly or by using the methods from \refsec{genvisrep}.
Then we will use this spatial layout to create a partitioning of the hypergraph in a desired number of $m \in \N$ parts.
To do so, we employ the \texttt{k-means++} method \cite{Arthur2007}, given by \refalg{kmeans}.
This algorithm searches for $m$ centers $z_1, \ldots, z_m \in \R^d$ such that
\begin{equation} \label{eq:kmeans}
\sum_{i = 1}^k \min_{1 \leq j \leq m} \|x_i - z_j\|^2
\end{equation}
is minimal, which is NP-hard for all $m \geq 2$ \cite{Aloise2009}.
The advantage of \texttt{k-means++} is that the algorithm finds $z_1, \ldots, z_m$ in $\mathcal{O}(k \, m \, \log m)$ time, while the value of \refeq{kmeans} is expected to be within a factor of $8 \, (\log m + 2)$ from its minimum value already at the start (\refalgline{kmeansstart}) of the first iteration \cite{Arthur2007}.
The \texttt{k-means++} algorithm therefore permits us to isolate clusters quickly in the visual representation of our hypergraph, which correspond to highly interconnected patches of vertices in the hypergraph.
\refAlg{kmeans} is furthermore easily parallelized in shared memory (lines \ref{algline:kmeanspar1} and \ref{algline:kmeanspar2}).
The sum at \refalgline{kmeanssum} can also be performed in parallel by computing partial sums for the $z_j$ and summing these afterwards.

\begin{algorithm}
\begin{algorithmic}[1]
\STATE set $z_1$ to a randomly chosen point from $\{x_1, \ldots, x_k\}$;
\FOR{$n = 2$ \TO $m$}
	\FOR{$i = 1$ \TO $k$ \textbf{parallel}} \label{algline:kmeanspar1}
		\STATE $d_i \gets \min_{1 \leq j < n} \|x_i - z_j\|^2$;
	\ENDFOR
	\STATE choose $z_n$ to be equal to $x_i$ with probability $d_i / (d_1 + \ldots + d_k)$;
\ENDFOR
\FOR{a fixed number of iterations} \label{algline:kmeansstart}
	\FOR{$i = 1$ \TO $k$ \textbf{parallel}} \label{algline:kmeanspar2}
		\STATE $j_i \gets \argmin_{1 \leq j \leq m} \|x_i - z_j\|^2$;
	\ENDFOR
	\FOR{$j = 1$ \TO $m$}
		\STATE $C_j := \{1 \leq i \leq k \setsep j_i = j\}$;
		\STATE $z_j := \frac{1}{|C_j|} \sum_{i \in C_j} x_i$; \label{algline:kmeanssum}
	\ENDFOR
\ENDFOR
\end{algorithmic}
\caption{The \texttt{k-means++} algorithm \cite{Arthur2007} finds centers $z_1, \ldots, z_m \in \R^d$ for a given set of points $x_1, \ldots, x_k \in \R^d$, trying to minimize \refeq{kmeans}.}
\label{alg:kmeans}
\end{algorithm}

The $m$ disjoint subsets $C_1, \ldots, C_m \subseteq V$ produced by \refalg{kmeans} form an $m$-way partitioning of $V$.
It should be remarked that this way of generating a partitioning does not enforce balancing of the partitioning, but in general \texttt{k-means++} does a good job of dividing the point set into parts of approximately equal size: large groups of points pull centers harder towards themselves, which enlarges the other, smaller, groups.
Such balancing can be observed in \reffig{vmoprocess} and \reffig{permute2}.
Partitionings generated by \refalg{kmeans} can further be improved by subjecting them to a few iterations of the Kernighan--Lin algorithm \cite{Kernighan1970}.

\section{LU decomposition} \label{sec:lu}

We will now apply the ideas discussed in the previous sections to performing an LU decomposition of a given matrix in parallel using nested dissection \cite{George1973, Hendrickson1998}, see also \cite{Aykanat2004, Catalyurek2009, Grigori2010, Hu2000, Mehrabi1993}.

\begin{definition} \label{def:lu}
Let $A \in \C^{m \times m}$ be a given $m \times m$ matrix.
Then a \emph{(permuted) LU decomposition} of $A$ is a decomposition of the form
\begin{equation} \label{eq:ludecomposition}
P \, A \, Q = L \, U,
\end{equation}
where $P, Q \in \{0, 1\}^{m \times m}$ are permutation matrices and $L, U \in \C^{m \times m}$ with entries $l_{i \, j}$, $u_{i \, j}$ respectively, such that $l_{i \, j} = 0$ for $i < j$, $u_{i \, j} = 0$ for $i > j$, and $l_{i \, i} = 1$ for all $i$.
\end{definition}

\begin{algorithm}
\begin{algorithmic}[1]
\FOR{$k = 1$ \TO $m$}
	\STATE $\pi_k \gets k$, $\sigma_k \gets k$;
\ENDFOR
\FOR{$k = 1$ \TO $m$} \label{algline:blockludiagonal}
	\STATE \label{algline:blocklupivot} find $k \leq i, j \leq m$ such that $|a_{i \, j}|$ is maximal;
	\STATE swap rows $k$ and $i$ and swap $\pi_k$ and $\pi_i$;
	\STATE swap columns $k$ and $j$ and swap $\sigma_k$ and $\sigma_j$;
	\IF{$a_{k \, k} \neq 0$}
		\FOR{$i = k + 1$ \TO $m$}
			\STATE $a_{i \, k} \gets a_{i \, k} / a_{k \, k}$;
		\ENDFOR
		\FOR{$i = k + 1$ \TO $m$}
			\FOR{$j = k + 1$ \TO $m$}
				\STATE $a_{i \, j} \gets a_{i \, j} - a_{i \, k} \, a_{k \, j}$; \label{algline:blockluinner}
			\ENDFOR
		\ENDFOR
	\ENDIF
\ENDFOR
\STATE set $p_{i \, j}$ to $1$ if $\pi_i = j$ and $0$ otherwise, to form $P$;
\STATE set $q_{i \, j}$ to $1$ if $\sigma_j = i$ and $0$ otherwise, to form $Q$;
\STATE set $l_{i \, j}$ to $a_{i \, j}$ if $i > j$, $1$ if $i = j$, and $0$ otherwise, to form $L$;
\STATE set $u_{i \, j}$ to $a_{i \, j}$ if $i \leq j$ and $0$ otherwise, to form $U$;
\end{algorithmic}
\caption{LU decomposition (algorithm 3.4.2 from \cite{Golub1996}).
Determines for a matrix $A \in \C^{m \times m}$ with entries $a_{i \, j}$ factors $P$, $Q$, $L$, and $U$ such that $P \, A \, Q = L \, U$.}
\label{alg:blocklu}
\end{algorithm}

We include permutations in \refdef{lu} to ensure that well-behaved matrices like
$\left( \begin{smallmatrix} 0 & 1 \\ 1 & 0 \end{smallmatrix} \right)$
also have an LU decomposition.
Calculating an LU decomposition for a given matrix with complete pivoting is described by \refalg{blocklu}.
While this algorithm is fine for small dense matrices, we get into trouble with large sparse matrices for two reasons: \refalg{blocklu} takes $\mathcal{O}(m^3)$ iterations (which scales rather badly), regardless of matrix sparsity, and as illustrated by \refexa{fillin}, the sparsity of the original matrix may be lost during the decomposition, requiring up to $\mathcal{O}(m^2)$ memory.
We will remedy these problems by calculating appropriate permutation matrices $P$ and $Q$, using the techniques from the previous sections.

\begin{example} \label{exa:fillin}
Consider the following decomposition (using \refalg{blocklu} without pivoting):
$$\left( \begin{matrix} 2 & 1 & 1 & 1 \\
 1 & 2 & 0 & 0 \\
 1 & 0 & 2 & 0 \\
 1 & 0 & 0 & 2 \end{matrix} \right)
\, = \, 
\left( \begin{matrix} 1 & 0 & 0 & 0 \\
 \frac{1}{2} & 1 & 0 & 0 \\
 \frac{1}{2} & -\frac{1}{3} & 1 & 0 \\
 \frac{1}{2} & -\frac{1}{3} & -\frac{1}{2} & 1 \end{matrix} \right) \,
\left( \begin{matrix} 2 & 1 & 1 & 1 \\
 0 & \frac{3}{2} & -\frac{1}{2} & -\frac{1}{2} \\
 0 & 0 & \frac{4}{3} & -\frac{2}{3} \\
 0 & 0 & 0 & 1 \end{matrix} \right).$$
Here, the sparsity pattern of the original matrix is lost completely in the $L$ and $U$ factors, and a \emph{fill-in} of six new nonzeros is created.
Suppose we permute the matrix by swapping the first and last rows and columns, then we obtain the following decomposition:
$$\left( \begin{matrix} 2 & 0 & 0 & 1 \\
 0 & 2 & 0 & 1 \\
 0 & 0 & 2 & 1 \\
 1 & 1 & 1 & 2 \end{matrix} \right)
\, = \, 
\left( \begin{matrix} 1 & 0 & 0 & 0 \\
 0 & 1 & 0 & 0 \\
 0 & 0 & 1 & 0 \\
 \frac{1}{2} & \frac{1}{2} & \frac{1}{2} & 1 \end{matrix} \right) \,
\left( \begin{matrix} 2 & 0 & 0 & 1 \\
 0 & 2 & 0 & 1 \\
 0 & 0 & 2 & 1 \\
 0 & 0 & 0 & \frac{1}{2} \end{matrix} \right).$$
Now, $L$ and $U$ have the same sparsity pattern as the original matrix.
Furthermore, performing \refalg{blocklu} on the permuted matrix also required less work because of zeros that were preserved during decomposition: the two loops around \refalgline{blockluinner} can skip most rows and columns because either $a_{i \, k}$ or $a_{k \, j}$ is equal to $0$.
\end{example}

\subsection{Recursive BBD form} \label{sec:recbbd}

\begin{figure}
\begin{center}
\includegraphics[width=9cm]{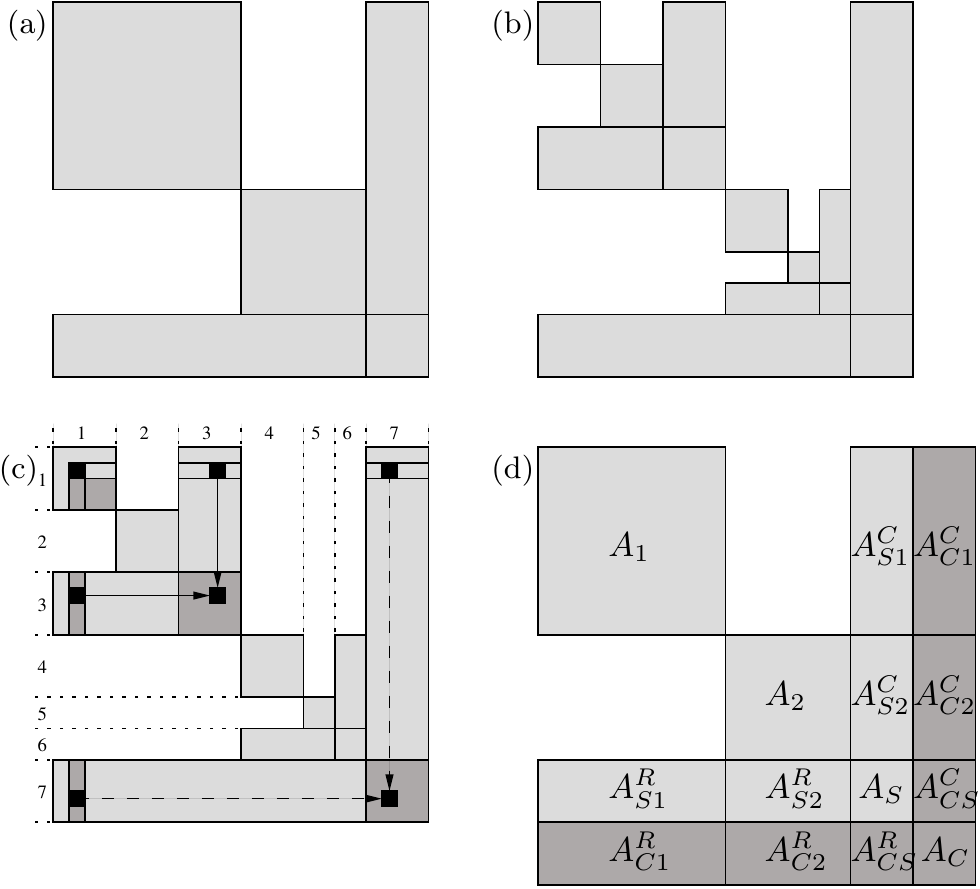}
\end{center}
\caption{Nested dissection for LU decomposition. (a) Bordered Block Diagonal (BBD) matrix form; (b) recursive BBD matrix form; (c) LU decomposition contributions; (d) matrix form for \refalg{parlu}.}
\label{fig:recursivebbd}
\end{figure}

Following \refexa{fillin} we will try to bring the sparse matrix into Bordered Block Diagonal (BBD) form \cite{Hu2000} as illustrated in \reffig{recursivebbd} (a).
The block in the lower-right corner is commonly called the \emph{Schur complement}.
A row that contains a nonzero in a column that intersects the first diagonal block and also in a column that intersects the second diagonal block is said to be \emph{cut} or \emph{split} with respect to the subdivision of the matrix.
Cut columns are defined similarly.
Performing \refalg{blocklu} on a matrix in such a form will only generate fill-in in the shaded blocks and the costly inner loop can skip the empty blocks.
By doing this recursively, see \reffig{recursivebbd} (b), the amount of fill-in will be further reduced.
This principle is called \emph{nested dissection} \cite{George1973}.

Such a recursive matrix layout furthermore permits us to create a parallel LU decomposition algorithm (similar to \cite{Mehrabi1993}).
We illustrate this in \reffig{recursivebbd} (c) where we are busy performing \refalg{blocklu} along the diagonal of diagonal block 1 (for the sake of simplicity we do not perform any pivoting).
For this nonzero on the diagonal, performing LU decomposition will only modify the darkly shaded parts of the matrix and therefore leave the diagonal blocks 2, 4, and 5 untouched.
Furthermore, the LU decomposition contributions of all diagonal blocks to the Schur complements 3, 6, and 7 do not depend on the actual values in the Schur complements, so we can perform LU decomposition on blocks 1, 2, 4, and 5 in parallel and add the contributions to the Schur complements afterwards.
To process the nonzero on the diagonal of block 1, we need nonzero values not only from the cut rows and columns of Schur complement 3 (solid arrows), but \emph{also} from the rows and columns of Schur complement 7 (dashed arrows).
So we need to keep track of \emph{all} previous Schur complements during parallel LU decomposition.

For the parallel LU algorithm outlined in \refalg{parlu}, we therefore work recursively on a matrix of the form illustrated in \reffig{recursivebbd} (d).
Here, we added \emph{contribution blocks} $A_C$, $A^C_{C1}$, $A^C_{C2}$, $A^C_{CS}$, $A^R_{C1}$, $A^R_{C2}$, and $A^R_{CS}$ to the matrix, which are indicated by a darker shade.
At the start of \refalg{parlu} (so for the original matrix $A$), these are empty, but as the algorithm further recurses, the contribution blocks will contain all the contributions of the LU decomposition to Schur complements of the previous recursion levels (i.e. the data required for the dashed arrows in \reffig{recursivebbd} (c)).
In an implementation of \refalg{parlu} it would be efficient to store the nonzeros $a_{i \, j}$ of the matrix by increasing $\min \{i, j\}$.
Thus, we keep all data necessary to perform the LU decomposition at \refalgline{parlublock2} together, regardless of the level of recursion.
This is better than storing the nonzeros by increasing $i$ (Compressed Row Storage, CRS) or $j$.

\begin{algorithm}
\begin{algorithmic}[1]
\IF{we wish to continue subdividing} \label{algline:parlusubdivide}
	\STATE apply \refalg{parlu} recursively and in parallel to
$$\includegraphics[width=7cm]{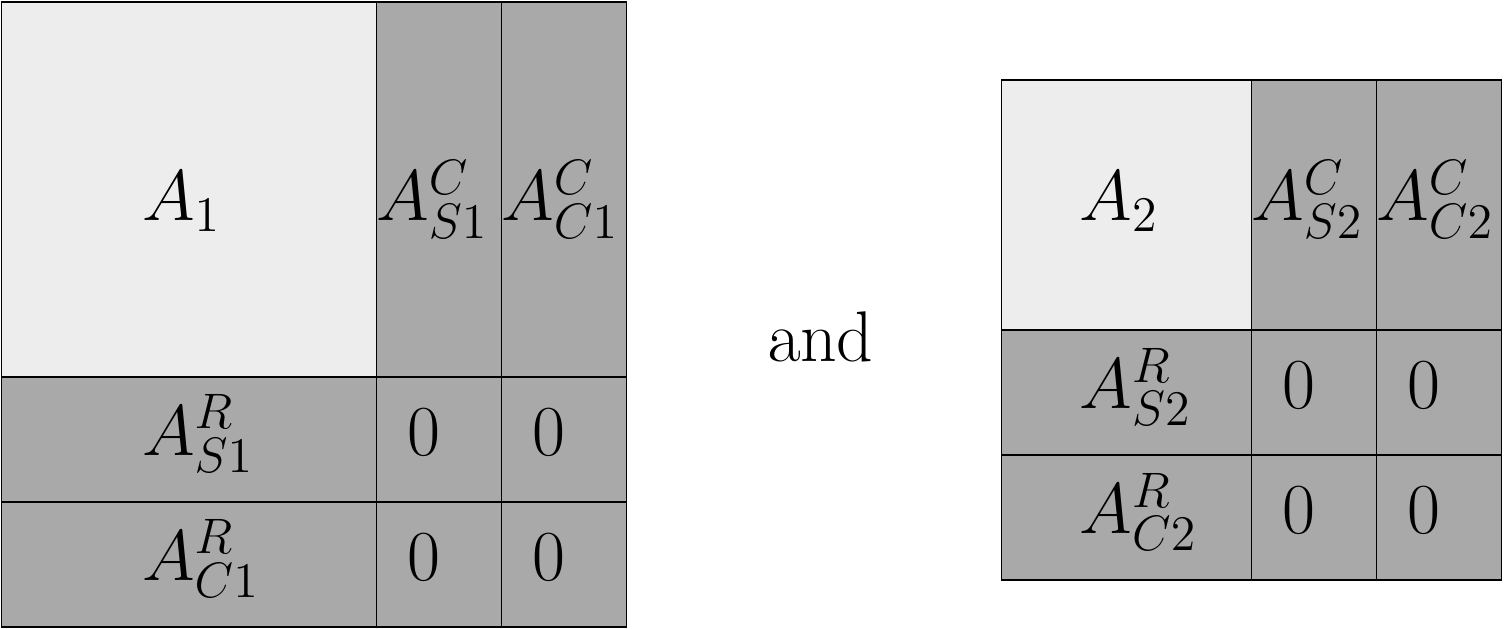}$$
	\STATE add the contributions from $A_1$ and $A_2$ to $A_S$, $A_C$, $A^R_{CS}$, and $A^C_{CS}$;
	\STATE \label{algline:parlublock1} perform LU decomposition (e.g. \refalg{blocklu}) on 
$$\includegraphics[width=1.2cm]{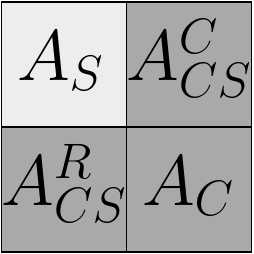},$$
only permuting within $A_S$ and stopping after factorizing $A_S$;
\ELSE
	\STATE \label{algline:parlublock2} perform LU decomposition on $A$, only permuting within the lightly shaded part of the matrix and stopping after factorizing $A_1$, $A_2$, and $A_S$;
\ENDIF
\end{algorithmic}
\caption{Parallel recursive LU decomposition of the matrix from \reffig{recursivebbd} (d).}
\label{alg:parlu}
\end{algorithm}

It is important to note that the LU decompositions performed in \refalg{parlu} on \refalgline{parlublock1} and \refalgline{parlublock2} are incomplete in the sense that they only treat part of the given matrix.
In \refalg{blocklu}, this would amount to specifying a number $1 \leq m' < m$ and letting $k$ run from $1$ to $m'$ at \refalgline{blockludiagonal} and choosing $i$ and $j$ such that $k \leq i, j \leq m'$ at \refalgline{blocklupivot}.
To improve performance, multifrontal \cite{Davis1997} or supernodal \cite{Demmel1999} methods could be used to perform these LU decompositions.
The condition at \refalgline{parlusubdivide} in \refalg{parlu} is used to stop the algorithm whenever the resulting diagonal block becomes so small that a direct LU decomposition would outperform further recursion, or when there is a risk of the diagonal matrices becoming singular.
Note that \refalg{parlu} is \emph{processor-oblivious} in the sense that we can continue recursing on the diagonal blocks while there are still more processors available, up to the recursion depth where the diagonal blocks are still sufficiently large.

Since with this parallel method we can only pivot within each block (not doing so would destroy the recursive BBD form), we could encounter a singular submatrix, as illustrated by \refthm{luranks} and \refexa{luproblem}.

\begin{theorem} \label{thm:luranks}
Let $A \in \C^{a \times a}$, $B \in \C^{b \times b}$, $C \in \C^{c \times c}$ such that $d := a - (b + c) \geq 0$ and $A$ is of the form
$$A = \left( \begin{array}{ccc}
\cline{1-1} \cline{3-3}
\multicolumn{1}{|c|}{B} & 0 & \multicolumn{1}{|c|}{} \\
\cline{1-2}
0 & \multicolumn{1}{|c|}{C} & \multicolumn{1}{|c|}{} \\
\cline{1-3}
\multicolumn{2}{|c|}{} & \multicolumn{1}{|c|}{D} \\
\cline{1-3}
\end{array} \right).$$

If $\det(A) \neq 0$, then
\begin{equation} \label{eq:luranks}
b + c - d \leq \rank(B) + \rank(C) \leq b + c.
\end{equation}
\end{theorem}
\begin{proof}
First of all, note that if the matrix $A' \in \C^{a \times a}$ is obtained from $A$ using Gauss--Jordan elimination with column pivoting, then $\det(A) = 0$ if and only if $\det(A') = 0$.
Suppose that $\det(A) \neq 0$, then by performing these operations on $B$ and $C$ separately, we find the nonzero value
\begin{align*}
& \det \left( \begin{array}{ccccc}
\cline{1-2} \cline{5-5}
\multicolumn{1}{|c|}{I_{\rank(B)}} & \multicolumn{1}{|c|}{} & 0 & 0 & \multicolumn{1}{|c|}{} \\
\cline{1-2}
0 & 0 & 0 & 0 & \multicolumn{1}{|c|}{} \\
\cline{3-4}
0 & 0 & \multicolumn{1}{|c|}{I_{\rank(C)}} & \multicolumn{1}{|c|}{} & \multicolumn{1}{|c|}{} \\
\cline{3-4}
0 & 0 & 0 & 0 & \multicolumn{1}{|c|}{} \\
\cline{2-2} \cline{4-5}
0 & \multicolumn{1}{|c|}{} & 0 & \multicolumn{1}{|c|}{} & \multicolumn{1}{|c|}{D} \\
\cline{2-2} \cline{4-5}
\end{array} \right)
 = \pm \det \left( \begin{array}{ccccc}
\cline{1-1} \cline{4-5}
\multicolumn{1}{|c|}{I_{\rank(B)}} & 0 & 0 & \multicolumn{1}{|c|}{} & \multicolumn{1}{|c|}{} \\
\cline{1-3} \cline{4-5}
0 & \multicolumn{1}{|c|}{I_{\rank(C)}} & \multicolumn{1}{|c|}{} & 0 & \multicolumn{1}{|c|}{} \\
\cline{2-3}
0 & 0 & 0 & 0 & \multicolumn{1}{|c|}{} \\
0 & 0 & 0 & 0 & \multicolumn{1}{|c|}{} \\
\cline{3-5}
0 & 0 & \multicolumn{1}{|c|}{} & \multicolumn{1}{|c|}{} & \multicolumn{1}{|c|}{D} \\
\cline{3-5}
\end{array} \right)
\\
& = \pm \det \left( \begin{array}{ccc}
\cline{3-3}
 0 & 0 & \multicolumn{1}{|c|}{} \\
 0 & 0 & \multicolumn{1}{|c|}{} \\
\cline{1-3}
\multicolumn{1}{|c|}{} & \multicolumn{1}{|c|}{} & \multicolumn{1}{|c|}{D} \\
\cline{1-3}
\end{array} \right).
\end{align*}
The resulting smaller  matrix has size $a - \rank(B) - \rank(C)$ and must be of maximum rank because its determinant is nonzero.
Let $e := a - \rank(B) - \rank(C) - d$, then $e \geq a - b - c - d = 0$.
If $e \leq d$, a matrix with the above nonzero pattern can have maximum rank $e + d$.
If $e > d$, the rank of such a matrix can be at most $2 \, d < e + d$.
Therefore, it is necessary that $e \leq d \iff a - \rank(B) - \rank(C) - d \leq d \iff a - 2 \, d \leq \rank(B) + \rank(C) \iff (b + c + d) - 2 \, d \leq \rank(B) + \rank(C)$, from which \refeq{luranks} follows.
\end{proof}

\refThm{luranks} shows us that we cannot assume our diagonal blocks to be invertible whenever the Schur complement is nonempty.
Furthermore, it motivates us to reduce the size of the Schur complement: this will increase the minimum rank that the diagonal blocks are required to have and thereby increases stability.
In terms of hypergraph partitioning we therefore see that we should at all times try to make the Schur complement \emph{as small as possible}: this will increase parallelism in the sense that more rows/columns can be treated in parallel by \refalg{parlu}, it will reduce fill-in, and it will improve stability.\footnote{So hypergraph partitioners used for the purpose of bringing the matrix into recursive BBD form should use the \emph{cut-net} metric instead of the $(\lambda - 1)$-metric, reducing the number of cut hyperedges, and not the associated communication volume.}

To prevent the diagonal blocks from becoming singular we allow for an optional specification of a desired (strengthened) matrix diagonal beforehand \cite{Duff2001}, which will be preserved by the generated permutations as described in \refsec{permutations}.
As \refexa{luproblem} shows however, this is not guaranteed to solve the problem.

\begin{example} \label{exa:luproblem}
Let
$$A = \left( \begin{matrix}
 2 & 1 & 0 & 0 & 1 \\
 4 & 2 & 0 & 0 & 1 \\
 0 & 0 & 2 & 1 & 1 \\
 0 & 0 & 1 & 2 & 1 \\
 1 & 1 & 1 & 1 & 2
\end{matrix} \right), \quad
B = \left( \begin{matrix}
 2 & 1 \\
 4 & 2
\end{matrix} \right), \quad
C = \left( \begin{matrix}
 2 & 1 \\
 1 & 2
\end{matrix} \right).$$
Then $a = 5$, $b = 2$, $c = 2$, $d = 1$, $\rank(B) = 1$, $\rank(C) = 2$, and $\det(A) = 3 \neq 0$.
Therefore, the bound in \refeq{luranks} is tight: $b + c - d = 3 = \rank(B) + \rank(C)$.
Note that $\det(B) = 0$ even though the product of the diagonal elements of $A$ is maximal.
\end{example}

During the performed benchmark with SuperLU (\reftab{superlucomparison}) we found that for $8$ of the $28$ matrices no pivoting was required at all (not even in the Schur complements), and in all other cases pivoting with a threshold of $10^{-6}$ was sufficient.
Therefore, if we keep the Schur complements small, strengthen the matrix diagonal, and use threshold pivoting within diagonal blocks and Schur complements, we anticipate that submatrix singularity will not pose any significant problems in practice.

\subsection{Permutations} \label{sec:permutations}

We will now apply the ideas from sections \ref{sec:hypergraphs}, \ref{sec:genvisrep}, and \ref{sec:partitioning} to obtain the desired permutations to bring a given sparse matrix $A \in \C^{m \times m}$ with $\mathit{nz}$ nonzeros into recursive BBD form.
This method will be referred to as \emph{visual matrix ordering} (VMO).
We assume the matrix to be square, because we want to use VMO for LU decomposition.

Firstly, we need to determine what kind of hypergraph we will use to represent $A$ (from \reftab{hypergraphmatrix}).
Using only the symmetric representation is not appropriate, because LU factorization is usually applied to unsymmetric matrices.
The column-net and row-net approach often do not yield optimal partitionings when compared to the finegrain representation \cite{Catalyurek2001}.
However, the finegrain representation results in $\mathit{nz}$ vertices, thus degrading the performance of \refalg{vmominenergy}, which would scale as $\mathcal{O}(\mathit{nz} \, \log(\mathit{nz}) + m)$.
Inspection of the visual layouts revealed that a good layout for the finegrain representation could be obtained by laying out its dual (the bipartite representation, which is a graph) and then mapping each nonzero to the average of the points of the row and column belonging to that particular nonzero.
As the bipartite representation has only $2 \, m$ vertices instead of $\mathit{nz}$, the layout can be generated much faster, scaling as $\mathcal{O}(m \, \log(m) + \mathit{nz})$.
It also permits us easily to maintain a previously selected strengthened diagonal in the generated permutations

Therefore, let $G = (V, E)$ be the bipartite representation of our sparse matrix $A$.
To avoid the problem of ending up with a singular matrix during recursion, we permit a desired strengthened diagonal to be specified with the matrix, in the form of a perfect bipartite graph matching $M \subseteq E$, \cite{Duff2001}.
We will view this matching as a map $\mu : V \rightarrow V$ which maps each vertex $v \in V$ to $\mu(v) \in V$ such that the edge $\{v, \mu(v)\} \in M$ (as $M$ is a perfect matching, exactly one vertex $\mu(v)$ has this property).

We can also incorporate the values of the nonzeros of the matrix in the partitioning by setting the edge costs of $G$ to $|a_{i \, j}|$ for each edge $\{i, j\} \in E$ (optionally rescaling these values to a fixed interval to avoid convergence issues in \refalg{vmominenergy}).
This is natural, because zeros of the matrix are not incorporated at all in \refeq{graphenergy} (as they are not included in $E$), so letting the edge cost of $\{i, j\}$ go to $0$ as $|a_{i \, j}| \rightarrow 0$ gradually decreases the influence of $\{i, j\}$ on the energy function of $G$ to zero.
However, as this reduced the quality of the partitionings in terms of fill-in, we did not use this option for the performed experiments.

\begin{algorithm}
\begin{algorithmic}[1]
\STATE determine two centers $z_1, z_2 \in \R^d$ in $x(V) \subseteq \R^d$ by \refalg{kmeans};
\FORALL{$v \in V$ in parallel}
	\IF{$\|x(v) - z_1\| \leq \|x(v) - z_2\|$}
		\STATE $p(v) \gets 1$;
	\ELSE
		\STATE $p(v) \gets 2$;
	\ENDIF
\ENDFOR
\STATE $r \gets z_2 - z_1$; $\delta \gets \frac{1}{2} (z_2 + z_1) \cdot r$; \label{algline:permuteplane}
\FORALL{$e = \{v, w\} \in E$} \label{algline:vertexseparator}
	\IF{$\{p(v), p(w)\} = \{1, 2\}$}
		\IF{$|x(v) \cdot r - \delta| \leq |x(w) \cdot r - \delta|$}
			\STATE $p(v) \gets 3$; $p(\mu(v)) \gets 3$; \label{algline:respectmatch1}
		\ELSE
			\STATE $p(w) \gets 3$; $p(\mu(w)) \gets 3$; \label{algline:respectmatch2}
		\ENDIF
	\ENDIF
\ENDFOR
\FORALL{$e = \{v, w\} \in E$ in parallel}
	\STATE $q(e) \gets \max \{p(v), p(w)\}$;
\ENDFOR
\STATE sort the vertex pairs $\{v, \mu(v)\}$ by their $p$-values to obtain $V_1, V_2, V_3$; \label{algline:permutecountsort}
\STATE sort the edges by their $q$-values to obtain $E_1, E_2, E_3$;
\end{algorithmic}
\caption{Given a graph $G = (V, E)$ with visual representation $x : V \rightarrow \R^d$ and a map $\mu : V \rightarrow V$ derived from a perfect bipartite graph matching $M \subseteq E$, this algorithm partitions $V$ and $E$ into $V_1, V_2, V_3$ and $E_1, E_2, E_3$ respectively,
where no $\{v, w\} \in E$ exists with $v \in V_1$ and $w \in V_2$, and such that $e \in E_1 \rightarrow e \subseteq V_1$, $e \in E_2 \rightarrow e \subseteq V_2$, and $e \in E_3 \rightarrow e \cap V_3 \neq \emptyset$ (\reffig{permute} (left)).}
\label{alg:vmopermute}
\end{algorithm}

\begin{figure}
\begin{center}
\begin{tabular}{c@{\hspace{0.5cm}}c@{\hspace{0.5cm}}c}
\includegraphics[width=4cm]{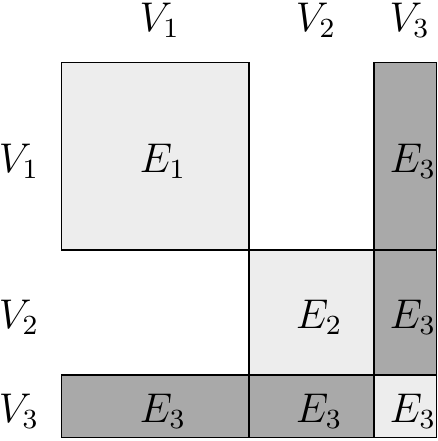} &
\includegraphics[width=3.5cm]{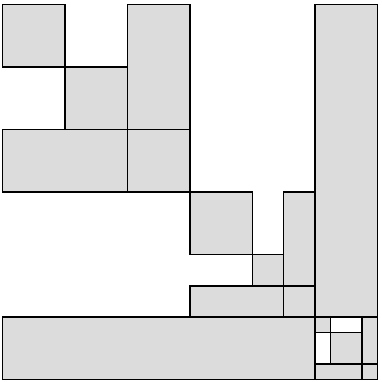} &
\includegraphics[width=3.5cm]{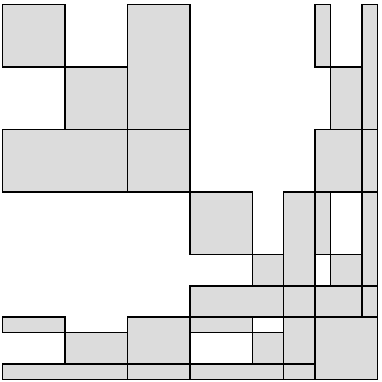}
\end{tabular}
\end{center}
\caption{Matrix partitioning for \refalg{vmopermute} (left) and improved partitionings obtained by bringing either the Schur complement (middle) or the cut rows and columns (right) into BBD form.}
\label{fig:permute}
\end{figure}

Using \refalg{vmovisrep} we generate a visual representation of $G$.
Then we apply \refalg{vmopermute} to obtain $V = V_1 \cup V_2 \cup V_3$ and $E = E_1 \cup E_2 \cup E_3$, where $p(v)$ and $q(e)$ denote the part indices for vertices $v$ and edges $e$.
\refAlg{vmopermute} turns the edge separator, obtained by partitioning the vertices using \refalg{kmeans}, into a vertex separator.
To do so we exploit the geometry of the partitioning by letting the vertices closest to the plane (described in \refalg{vmopermute} by its normal $r$ and distance $\delta$ on \refalgline{permuteplane}) separating the two groups of vertices be chosen to be added to the vertex separator.
We ensure that we preserve the strengthened diagonal in lines \ref{algline:respectmatch1} and \ref{algline:respectmatch2}: this ensures that edges from the matching $M$ are contained completely in either $V_1$, $V_2$, or $V_3$, which prevents them from entering the darker off-diagonal blocks in \reffig{permute}.
This can be skipped if no matching is available or desired (e.g. in the context of sparse matrix--vector multiplication instead of LU decomposition), which will result in smaller $V_3$ and $E_3$.
However, for LU decomposition it is necessary, in particular to maintain square blocks on the diagonal.
After the first iteration of \refalg{vmopermute}, we again apply it to $G_1 = (V_1, E_1)$ and $G_2 = (V_2, E_2)$ and continue doing this recursively to obtain recursive BBD permutations for our matrix $A$ as shown in the rightmost column of \reffig{vmoprocess}.

The permutations themselves can directly be obtained from the recursive partitioning of $V$: the rows and columns of the block $E_1$ (see \reffig{permute}) are exactly the vertices in $V_1$, and similarly for the rows and columns of the blocks $E_2$ and $E_3$.
Therefore, a simple linear walk through the reordered vertices (\refalgline{permutecountsort} of \refalg{vmopermute}) will provide the proper permutations of the rows and columns of our matrix $A$.

When permuting the matrix to recursive BBD form, we have additional freedom in permuting the rows and columns of $V_3$ in \reffig{permute} (left) (also see \reffig{permute2}).
A direct way to do this is also to apply \refalg{vmopermute} recursively to $G_3 = (V_3, E'_3)$, just like we do for $G_1$ and $G_2$.
Here $E'_3$ consists of all edges $e \in E_3$ satisfying $e \subseteq V_3$ (so $E'_3$ is the lightly shaded $E_3$ part of \reffig{permute} (left)).
This gives permutations as illustrated in \reffig{permute} (middle).
An advantage of this method is that the strengthened diagonal is also maintained within the Schur complement.

Another way in which the additional freedom can be used, is to bring the cut rows and columns into recursive BBD form as illustrated in \reffig{permute} (right).
Doing this is a little more tricky: first of all, we assign a two-bit number to each vertex in $V_3$, initially set to $00$.
We also keep track of the edges $E_{1 \, 3}$ between $V_1$ and $V_3$, and edges $E_{2 \, 3}$ between $V_2$ and $V_3$.
Now, if $V_i$ ($i = 1, 2$) is split with \refalg{vmopermute} into $V_{i \, 1}$, $V_{i \, 2}$, and $V_{i \, 3}$ we can loop through all edges $\{v, w\}$ in $E_{i \, 3}$ with $w \in V_3$.
Then if $v \in V_{i \, 1}$ we set the first bit of the number associated with $w$ and if $v \in V_{i \, 2}$ we set the second bit.
If we do this for both splits of $V_1$ and $V_2$, and then sort the vertices in $V_3$ by their two-bit numbers we obtain a permutation as shown in \reffig{permute} (right).
By expanding these numbers to $b$ two-bit pairs and keeping track of the edges extending to the Schur complements for up to $b$ splits, we can bring the cut rows and columns into recursive BBD form up to the $b$th level.

\section{Experiments} \label{sec:experiments}

We implemented the VMO algorithm in C++ using the Intel Threading Building Blocks library for many-core parallelism where we chose to generate visual representations in $d = 4$ dimensions to be able to perform all parallel vector calculations in \refalg{vmominenergy} and \refalg{kmeans} efficiently on either the CPU (one Streaming SIMD \texttt{xmm*} register for a point in $\R^4$) or the GPU (a \texttt{float4} register for a point in $\R^4$).
This furthermore ensures that we do not need to take a square root in \refeq{graphenergygradientsimple}.

\begin{table}[ht]
\begin{center}
\begin{tabular}{|l|r|r|r|r|r|r|r|}
\hline
Matrix & Size & Nonzeros & VMO 			& CMD & MMD$+$ & MMD$\times$ \\
\hline
\texttt{swang1} & 3169 & 20841 			& \textbf{6.2} & 7.7 & 6.7 & 8.2 \\
\texttt{lns\_3937} & 3937 & 25407 		& \textbf{15.0} & 17.5 & 132.2 & 17.5 \\
\texttt{poli\_large} & 15575 & 33074 		& \textbf{1.6} & 1.6 & 1.6 & 1.6 \\
\texttt{mark3jac020}$^*$ & 9129 & 56175 	& 68.1 & 45.6 & 121.3 & \textbf{43.9} \\
\texttt{fd18}$^*$ & 16428 & 63406 		& \textbf{21.9} & 24.1 & 302.0 & 25.5 \\
\texttt{lhr04}$^*$ & 4101 & 82682 		& 6.0 & \textbf{4.1} & 20.6 & 4.3 \\
\texttt{raefsky6} & 3402 & 137845 		& \textbf{2.7} & 3.4 & 4.5 & 3.1 \\
\texttt{shermanACb}$^*$ & 18510 & 145149 	& 19.0 & 45.3 & \textbf{14.3} & 57.2 \\
\texttt{bayer04}$^*$ & 20545 & 159082 		& 10.2 & 4.2 & 41.8 & \textbf{4.2} \\
\texttt{Zhao2}$^*$ & 33861 & 166453 		& 158.1 & 115.1 & 1280.1 & \textbf{107.0} \\
\texttt{mult\_dcop\_03} & 25187 & 193216 	& 3.1 & \textbf{2.0} & 3.4 & 5.9 \\
\texttt{jan99jac120sc}$^*$ & 41374 & 260202 	& 71.8 & \textbf{15.9} & 52.4 & 19.7 \\
\texttt{bayer01}$^*$ & 57735 & 277774 		& 7.5 & \textbf{5.4} & 47.6 & 5.6 \\
\texttt{sinc12}$^*$ & 7500 & 294986 		& 37.8 & 44.7 & \textbf{36.3} & 45.3 \\
\texttt{onetone1}$^*$ & 36057 & 341088 		& 32.1 & 14.4 & 149.0 & \textbf{14.2} \\
\texttt{mark3jac140sc}$^*$ & 64089 & 399735 	& \textbf{111.0} & 125.7 & 4435.0 & 152.0 \\
\texttt{af23560} & 23560 & 484256 		& \textbf{24.8} & 25.0 & 82.7 & 26.9 \\
\texttt{e40r0100}$^*$ & 17281 & 553562 		& 9.2 & 9.2 & 137.5 & \textbf{8.4} \\
\texttt{sinc15}$^*$ & 11532 & 568526 		& 56.3 & 58.0 & \textbf{48.7} & 57.2 \\
\texttt{Zd\_Jac2\_db}$^*$ & 22835 & 676439 	& 9.6 & \textbf{5.1} & 32.1 & 5.7 \\
\texttt{lhr34c}$^*$ & 35152 & 764014 		& 7.0 & 4.7 & 50.5 & \textbf{4.7} \\
\texttt{sinc18}$^*$ & 16428 & 973826 		& \textbf{65.7} & 67.8 & 68.2 & 72.3 \\
\texttt{torso2} & 115967 & 1033473 		& 10.2 & 16.8 & \textbf{8.2} & 14.5 \\
\texttt{twotone} & 120750 & 1224224 		& 35.8 & \textbf{15.2} & 1448.1 & 17.0 \\
\texttt{lhr71c}$^*$ & 70304 & 1528092 		& 6.7 & 4.8 & 66.4 & \textbf{4.7} \\
\texttt{av41092}$^*$ & 41092 & 1683902 		& 64.6 & 26.0 & 177.6 & \textbf{23.8} \\
\texttt{bbmat}$^*$ & 38744 & 1771722 		& 32.0 & \textbf{26.7} & 1000.6 & 26.8 \\
\hline
\end{tabular}
\end{center}
\caption{Comparison between VMO and SuperLU 4.1 in terms of fill-in, defined as $(\mathit{nz}(L) + \mathit{nz}(U) - \mathit{nz}(I))/\mathit{nz}(A)$ for $A = L \, U$.
The best result for each matrix is \textbf{bold}, CMD = COLAMD, MMD$+$ = MMD($A^T + A$), and MMD$\times$ = MMD($A^T \, A$) are the column pre-orderings determined by SuperLU.
Matrices marked with $^*$ required threshold $10^{-6}$ pivoting for VMO.}
\label{tab:superlucomparison}
\end{table}

\begin{figure}[ht]
\begin{center}
\begin{tabular}{c@{\hspace{1cm}}c}
\includegraphics[width=5.0cm]{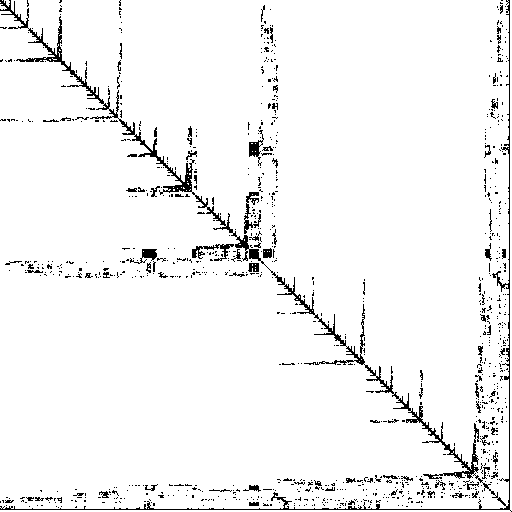} &
\includegraphics[width=5.0cm]{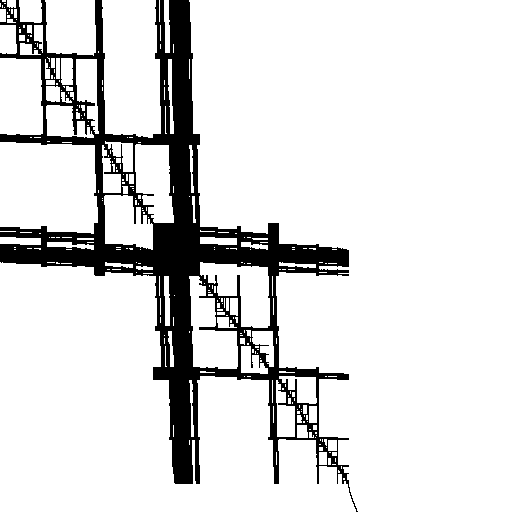}
\end{tabular}
\end{center}
\caption{The matrices \texttt{rhpentium} (left) and \texttt{wikipedia-20070206} (right), permuted by VMO to recursive BBD and recursive SBD form, respectively.
Additional permutation freedom is used for \texttt{rhpentium} as in \reffig{permute} (middle), and for \texttt{wikipedia-20070206} as in \reffig{permute} (right).}
\label{fig:permute2}
\end{figure}

To measure the quality of the generated permutations we compared VMO to the SuperLU \cite{Demmel1999} LU decomposition package by measuring fill-in, see \reftab{superlucomparison}.
In this case we made use of the additional freedom in the cut rows and columns by also recursively subdividing the cut parts of the graph while retaining the strengthened diagonal (\reffig{permute} (middle) and \reffig{permute2} (left)) to ensure that few small pivots are encountered along the diagonal.
We performed four decompositions for each matrix where we used permutations generated by VMO, as well as the built-in COLAMD($A$), MMD($A^T + A$), and MMD($A^T \, A$) column permutations generated by SuperLU.
For the permutations generated by SuperLU we retained the default SuperLU 4.1 options, while for the VMO permutations we first performed a run without any pivoting and then a run with threshold pivoting\footnote{SuperLU performs row pivoting by generating a row permutation $\pi$ such that for all $1 \leq j \leq m$, $|a_{\pi(j) \, j}| \geq u \, \max_{1 \leq i \leq m} |a_{i \, j}|$ where $u \in [0, 1]$ is the desired threshold, see \cite[eqn (4.4.7)]{Duff1986}.}, using a value of $u = 10^{-6}$.
To ensure we would not run into numerical problems we used a strengthened diagonal obtained via a heavy edge matching in the bipartite representation of $A$, augmented to a perfect matching via the Hopcroft--Karp algorithm \cite{Hopcroft1973}.
We furthermore validated the decomposition by comparing calculated condition numbers for all permutation methods and letting SuperLU calculate the backward error of the solution to $A \, x = b$ obtained by solving the system using the decomposition $A = L \, U$ (section 3.1 of \cite{Golub1996}) for $b = A \, (1, \ldots, 1)^T$.
\refTab{superlucomparison} shows that in $8$ of the $28$ cases, decomposition of the matrices permuted by VMO did not require \emph{any} pivoting at all, not even in the Schur complements.
From the table we see that VMO compares favorably with SuperLU:
looking at the lowest fill-in of COLAMD($A$), MMD($A^T + A$), and MMD($A^T \, A$) and the fill-in of VMO for each of the $28$ test matrices, we find that on average the fill-in of VMO equals $1.52$ times the lowest fill-in of the other methods, and that VMO outperforms all other methods in $8$ cases.
This indicates that the permutations generated by VMO are useful in the context of sparse LU decomposition.

We also compared VMO with Mondriaan \cite{Vastenhouw2005} in terms of matrix partitioning.
Firstly, we did this in the context of cache-oblivious sparse matrix--vector multiplication where the matrices are permuted into recursive Separated Block Diagonal (SBD) form \cite{Yzelman2009} (with the cut rows and columns in the middle instead of at the end) to decrease the number of cache-misses, independent of the particular cache hierarchy of the processor performing the multiplication.
Results are further improved by also using the additional freedom in the cut rows and columns to bring these into recursive SBD form (\reffig{permute2} (right)).
We measure the matrix multiplication time with the same program and on the same processor as \cite{Yzelman2010}: a single node of the Huygens supercomputer equipped with a dual-core 4.7GHz IBM Power6+ processor with 64kB L1 cache per core, a semi-shared L2 cache of 4MB, and an L3 cache of 32MB on which the matrix--vector multiplication program has been compiled with the IBM XL compiler.
In \reftab{spmvcomparison}, we compare the matrix--vector multiplication time for the original matrix with the best result from \cite{Yzelman2010} (where the matrix has been permuted by Mondriaan), and with the result obtained by using VMO.

\begin{table}[ht]
\begin{center}
\begin{tabular}{|l|r|r|r|r|r|r|}
\hline
Matrix & Rows & Columns & Nonzeros & Orig. & \cite{Yzelman2010} & VMO \\
\hline
\texttt{ex37} & 3565 & 3565 & 67591 & 0.116 & 0.113 & 0.113 \\
\texttt{memplus} & 17758 & 17758 & 126150 & 0.308 & 0.300 & 0.280 \\
\texttt{rhpentium} & 25187 & 25187 & 258265 & 0.645 & 0.627 & 0.646 \\
\texttt{lhr34} & 35152 & 35152 & 764014 & 1.37 & 1.34 & 1.34 \\
\texttt{lp\_nug30} & 52260 & 379350 & 1567800 & 5.35 & 4.85 & 9.15 \\
\texttt{s3dkt3m2} & 90449 & 90449 & 1921955 & 7.81 & 7.27 & 7.80 \\
\texttt{tbdlinux} & 112757 & 21067 & 2157675 & 6.43 & 5.03 & 5.66 \\
\texttt{stanford} & 281903 & 281903 & 2312497 & 19.0 & 9.35 & 5.88 \\
\texttt{stanford\_berkeley} & 683446 & 683446 & 7583376 & 20.9 & 19.2 & 22.5 \\
\texttt{wikipedia-20051105} & 1634989 & 1634989 & 19753078 & 249 & 116 & 128 \\
\texttt{cage14} & 1505785 & 1505785 & 27130349 & 69.4 & 74.4 & 99.0 \\
\texttt{wikipedia-20060925} & 2983494 & 2983494 & 37269096 & 688 & 256 & 264 \\
\hline
\end{tabular}
\end{center}
\caption{Comparison with Mondriaan in the context of cache-oblivious sparse matrix--vector multiplication \cite{Yzelman2010}.
We compare the original matrix--vector multiplication time with the best time from \cite{Yzelman2010} (which used Mondriaan 3.01 for reordering) and the best time with VMO.}
\label{tab:spmvcomparison}
\end{table}

VMO performs poorly for \texttt{lp\_nug30} and \texttt{cage14}.
For \texttt{lp\_nug30}, this can be explained by a lack of underlying geometrical structure: the visual representation of this matrix is a featureless blob from which little extra information can be obtained, resulting in quite bad permutations.
The matrix \texttt{cage14} already possesses a nonzero layout that is well suited for matrix--vector multiplication: both Mondriaan and VMO fail to improve the matrix--vector multiplication time.
For \texttt{tbdlinux}, \texttt{wikipedia-20051105}, and \texttt{wikipedia-20060925} VMO shows improvements comparable to those of Mondriaan, while for \texttt{memplus} and \texttt{stanford} the results are even better.
As generating the permutations with VMO is much faster (see \reftab{mondriaancomparison}), this makes VMO a viable alternative to Mondriaan in this context.

Another comparison with Mondriaan was made in terms of the cut-net metric, which is the appropriate metric in the context of LU decomposition because of \refthm{luranks}.
Hence, we look at the maximum number of cut rows and columns in all matrix (sub)divisions.
While Mondriaan divides the matrix among a given number of processors, VMO continues subdividing the matrix until it can no longer continue.
Therefore, we ran Mondriaan with a hybrid splitting strategy for the cut-net metric to divide the matrix into $256$ parts with a permitted imbalance of $0.1$ to obtain permutations comparable to those of VMO.

\begin{table}[ht]
\begin{center}
\begin{tabular}{|l|r|r|c|}
\hline
Matrix & Speedup & Speedup & Relative \\
 & \multicolumn{1}{c|}{V + O} & \multicolumn{1}{c|}{O} & cut \\
\hline
\texttt{ex37}				& 13.4 & 53.2 & 0.39 \\
\texttt{memplus}			& 2.1 & 11.0 & 2.86 \\
\texttt{rhpentium}			& 14.5 & 57.0 & 1.08 \\
\texttt{lhr34}				& 6.5 & 39.8 & 2.47 \\
\texttt{lp\_nug30}			& 9.1 & 144.7 & 2.20 \\
\texttt{s3dkt3m2}			& 4.7 & 26.7 & 1.02 \\
\texttt{tbdlinux}			& 25.0 & 228.3 & 0.75 \\
\texttt{stanford}			& 7.9 & 78.6 & 2.37 \\
\texttt{stanford\_berkeley}		& 17.0 & 118.0 & 3.26 \\
\texttt{wikipedia-20051105}		& 36.0 & 290.7 & 0.61 \\
\texttt{cage14}$^*$			& 4.3 & 29.1 & 0.62 \\
\texttt{wikipedia-20060925}$^*$ 	& 119.4 & 1104.3 & 1.02 \\
\hline
\end{tabular}
\end{center}
\caption{Comparison with Mondriaan in terms of the calculation time and the largest number of cut rows/columns in a split.
Speedup is defined as the time required by Mondriaan 3.01 to perform the matrix partitioning, divided by the time required by VMO to generate both the visual representation and permutations (V + O) or just the permutations (O).
The last column gives the maximum of the number of cut rows and columns in all splits of VMO, divided by the maximum obtained by Mondriaan.
Entries marked with a $^*$ were benchmarked on a different system because of a lack of memory.
The test matrices are the same as those from \reftab{spmvcomparison}.}
\label{tab:mondriaancomparison}
\end{table}

In \reftab{mondriaancomparison}, we measure the time it takes Mondriaan to perform the matrix partitioning and divide this by the time it takes VMO to generate a visual representation of the matrix and to generate a partitioning from this visual representation.
All timings except for those marked with $^*$ were measured on a system with a quad-core 2.8GHz Intel Core i7 860 processor and 8GB RAM, in particular to illustrate the gains of using VMO on a many-core system.
The entries marked with $^*$ needed to be benchmarked on a different system, because of Mondriaan's memory requirements: these were performed on a dual quad-core 2.4GHz AMD Opteron 2378 system with 32GB RAM.
From \reftab{mondriaancomparison} we find that VMO is on average $21.6$ times faster than Mondriaan, and if for all matrices a visual representation would already have been given the average speedup would even be $181.8$.
We also measure the maximum of the number of cut rows and columns in all subdivisions of the matrix for VMO and divide this by the maximum for Mondriaan.
This gives a measure for the relative maximum cut size when comparing the two methods: the maximum cut size obtained by VMO is on average $1.55$ times that of Mondriaan and in four cases it is less.
To make the comparison as fair as possible we used Mondriaan with the cut-net metric for partitioning, but it should still be remarked that minimizing the maximum number of cut rows and columns is not the primary objective of Mondriaan and the balancing restrictions placed on Mondriaan are absent for VMO.

\section{Conclusion}

We have shown that it is possible to create and use the visual representations of hypergraphs to generate partitionings and orderings which are of sufficient quality for sparse LU decomposition (\reftab{superlucomparison}) and sparse matrix--vector multiplication (\reftab{spmvcomparison}).
Our method generates orderings on average $21.6$ times faster than Mondriaan (\reftab{mondriaancomparison}).
We generalized the 2D/3D graph visualization method from \cite{Hu2005} to generate hypergraph geometries in higher dimensions.
Furthermore, the algorithms to generate visual representations (\refalg{vmominenergy}) and matrix orderings (\refalg{vmopermute}) are well suited to shared-memory many-core parallel architectures such as current many-core CPUs and GPUs.
We have implemented these algorithms in the software package VMO.

This also opens up opportunities for further research, such as moving from a shared-memory parallel architecture to distributed-memory, which would require significant modifications of \refalg{vmominenergy}, \refalg{vmopermute}, and the data structures involved.
Since VMO is fast and parallel, it also has potential to remove computational partitioning bottlenecks in large applications such as the human bone simulations in \cite{Bekas2010}.

\section*{Acknowledgments}

We thank Albert-Jan Yzelman for performing the sparse matrix--vector multiplication experiments (\reftab{spmvcomparison}).
We thank Job Kuit, Joop Kolk, and Paul Zegeling for helpful discussions and comments.
We thank the Dutch supercomputing center SARA in Amsterdam and the Netherlands National Computing Facilities foundation NCF for providing access to the Huygens supercomputer.

\bibliographystyle{siam}
\bibliography{refs}

\begin{thebibliography}{10}

\bibitem{Aloise2009}
{\sc D.~Aloise, A.~Deshpande, P.~Hansen, and P.~Popat}, {\em {NP}-hardness of
  {E}uclidean sum-of-squares clustering}, Machine Learning, 75 (2009),
  pp.~245--248.

\bibitem{Arthur2007}
{\sc D.~Arthur and S.~Vassilvitskii}, {\em k-means++: the advantages of careful
  seeding}, in SODA '07: Proceedings of the 18th annual ACM-SIAM symposium on
  Discrete algorithms, Philadelphia, PA, 2007, SIAM, pp.~1027--1035.

\bibitem{Axler1992}
{\sc S.~Axler, P.~Bourdon, and W.~Ramey}, {\em Harmonic function theory},
  vol.~137 of Graduate Texts in Mathematics, Springer-Verlag, New York, 1992.

\bibitem{Aykanat2004}
{\sc C.~Aykanat, A.~Pinar, and U.~V. \c{C}ataly\"{u}rek}, {\em Permuting sparse
  rectangular matrices into block-diagonal form}, SIAM J. Sci. Comput., 25
  (2004), pp.~1860--1879.

\bibitem{Bekas2010}
{\sc C.~Bekas, A.~Curioni, P.~Arbenz, C.~Flaig, G.~H. van Lenthe,
  R.~{M\"uller}, and A.~J. Wirth}, {\em Extreme scalability challenges in
  micro-finite element simulations of human bone}, Concurrency Computat.:
  Pract. Exper., 22 (2010), pp.~2282--2296.

\bibitem{Berge1976}
{\sc C.~Berge}, {\em Graphs and hypergraphs}, North-Holland, Amsterdam,
  revised~ed., 1976.

\bibitem{Catalyurek1999}
{\sc U.~V. \c{C}ataly\"urek and C.~Aykanat}, {\em Hypergraph-partitioning-based
  decomposition for parallel sparse-matrix vector multiplication}, IEEE Trans.
  Par. Dist. Syst., 10 (1999), pp.~673--693.

\bibitem{Catalyurek2001}
\leavevmode\vrule height 2pt depth -1.6pt width 23pt, {\em A fine-grain
  hypergraph model for {2D} decomposition of sparse matrices}, in Proceedings
  8th International Workshop on Solving Irregularly Structured Problems in
  Parallel, IEEE Press, Los Alamitos, CA, 2001, p.~118.

\bibitem{Catalyurek2009}
{\sc U.~V. \c{C}ataly\"urek, C.~Aykanat, and E.~Kayaaslan}, {\em Hypergraph
  partitioning-based fill-reducing ordering}, Technical Report
  OSUBMI-TR-2009-n02, Department of Biomedical Informatics, Ohio State
  University, Columbus, OH, April 2009.

\bibitem{Cormen2009}
{\sc T.~H. Cormen, C.~E. Leiserson, R.~L. Rivest, and C.~Stein}, {\em
  Introduction to algorithms}, MIT Press, Cambridge, MA, third~ed., 2009.

\bibitem{Davis1997}
{\sc T.~A. Davis and I.~S. Duff}, {\em An unsymmetric-pattern multifrontal
  method for sparse {LU} factorization}, SIAM J. Matrix Anal. Appl., 18 (1997),
  pp.~140--158.

\bibitem{Davis2010}
{\sc T.~A. Davis and Y.~F. Hu}, {\em The {U}niversity of {F}lorida sparse
  matrix collection}.
\newblock ACM Trans. Math. Software (to appear), 2010.

\bibitem{Demmel1999}
{\sc J.~W. Demmel, S.~C. Eisenstat, J.~R. Gilbert, X.~S. Li, and J.~W.~H. Liu},
  {\em A supernodal approach to sparse partial pivoting}, SIAM J. Matrix Anal.
  Appl., 20 (1999), pp.~720--755.

\bibitem{Duff1986}
{\sc I.~S. Duff, A.~M. Erisman, and J.~K. Reid}, {\em Direct Methods for Sparse
  Matrices}, Monographs on Numerical Analysis, Oxford University Press, Oxford,
  UK, 1986.

\bibitem{Duff2001}
{\sc I.~S. Duff and J.~Koster}, {\em On algorithms for permuting large entries
  to the diagonal of a sparse matrix}, SIAM J. Matrix Anal. Appl., 22 (2001),
  pp.~973--996.

\bibitem{Duistermaat2004i}
{\sc J.~J. Duistermaat and J.~A.~C. Kolk}, {\em Multidimensional Real Analysis
  {I}: Differentiation}, Cambridge University Press, Cambridge, UK, 2004.

\bibitem{Fruchterman1991}
{\sc T.~M.~J. Fruchterman and E.~M. Reingold}, {\em Graph drawing by
  force-directed placement}, Softw. Pract. Exper., 21 (1991), pp.~1129--1164.

\bibitem{George1973}
{\sc A.~George}, {\em Nested dissection of a regular finite element mesh}, SIAM
  J. Numer. Anal., 10 (1973), pp.~345--363.

\bibitem{Golub1996}
{\sc G.~H. Golub and C.~F.~Van Loan}, {\em Matrix Computations}, The Johns
  Hopkins University Press, 3rd~ed., 1996.

\bibitem{Grigori2010}
{\sc L.~Grigori, E.~G. Boman, S.~Donfack, and T.~A. Davis}, {\em
  Hypergraph-based unsymmetric nested dissection ordering for sparse {LU}
  factorization}, SIAM J. Sci. Comput., 32 (2010), pp.~3426--3446.

\bibitem{Hendrickson2000}
{\sc B.~Hendrickson and T.~G. Kolda}, {\em Graph partitioning models for
  parallel computing}, Parallel Comput., 26 (2000), pp.~1519--1534.

\bibitem{Hendrickson1998}
{\sc B.~Hendrickson and E.~Rothberg}, {\em Improving the run time and quality
  of nested dissection ordering}, SIAM J. Sci. Comput., 20 (1998),
  pp.~468--489.

\bibitem{Hopcroft1973}
{\sc J.~E. Hopcroft and R.~M. Karp}, {\em An {$n^{5/2}$} algorithm for maximum
  matchings in bipartite graphs}, SIAM J. Comput., 2 (1973), pp.~225--231.

\bibitem{Hu2005}
{\sc Y.~F. Hu}, {\em Efficient and high quality force-directed graph drawing},
  The Mathematica Journal, 10 (2005), pp.~37--71.

\bibitem{Hu2000}
{\sc Y.~F. Hu, K.~C.~F. Maguire, and R.~J. Blake}, {\em A multilevel
  unsymmetric matrix ordering algorithm for parallel process simulation},
  Comput. Chem. Engrg., 23 (2000), pp.~1631--1647.

\bibitem{Johnson1987}
{\sc C~Johnson}, {\em Numerical solution of partial differential equations by
  the finite-element method}, Cambridge University Press, Cambridge, UK, 1987.

\bibitem{Kernighan1970}
{\sc B.~W. Kernighan and S.~Lin}, {\em An efficient heuristic procedure for
  partitioning graphs}, Bell System Technical Journal, 49 (1970), pp.~291--307.

\bibitem{Mehrabi1993}
{\sc M.~R. Mehrabi and R.~A. Brown}, {\em An incomplete nested dissection
  algorithm for parallel direct solution of finite element discretizations of
  partial differential equations}, J. Sci. Comput., 8 (1993), pp.~373--387.

\bibitem{Nakasato2009}
{\sc N.~Nakasato}, {\em Oct-tree method on {GPU}: \$42/{G}flops cosmological
  simulation}.
\newblock arXiv:0909.0541v1 [astro-ph.IM], 2009.

\bibitem{Parter1961}
{\sc S.~Parter}, {\em The use of linear graphs in {G}auss elimination}, SIAM
  Rev., 3 (1961), pp.~119--130.

\bibitem{Vastenhouw2005}
{\sc B.~Vastenhouw and R.~H. Bisseling}, {\em A two-dimensional data
  distribution method for parallel sparse matrix-vector multiplication}, SIAM
  Rev., 47 (2005), pp.~67--95.

\bibitem{Walshaw2003}
{\sc C.~Walshaw}, {\em A multilevel algorithm for force-directed
  graph-drawing}, J. Graph Algorithms Appl., 7 (2003), pp.~253--285.

\bibitem{Yzelman2009}
{\sc A.~N. Yzelman and R.~H. Bisseling}, {\em Cache-oblivious sparse
  matrix--vector multiplication by using sparse matrix partitioning methods},
  SIAM J. Sci. Comput., 31 (2009), pp.~3128--3154.

\bibitem{Yzelman2010}
\leavevmode\vrule height 2pt depth -1.6pt width 23pt, {\em Two-dimensional
  cache-oblivious sparse matrix--vector multiplication}.
\newblock Preprint, 2010.

\end{thebibliography}

\end{document}